\documentclass{article}%
\usepackage{amsmath}
\usepackage{amsfonts}
\usepackage{amssymb}
\usepackage{graphicx}%
\providecommand{\U}[1]{\protect\rule{.1in}{.1in}}
\newtheorem{theorem}{Theorem}

\newtheorem{assumption}{Assumption}

\newtheorem{proposition}{Proposition}
\newtheorem{remark}{Remark}

\newenvironment{proof}[1][Proof]{\noindent\textbf{#1.} }{\ \rule{0.5em}{0.5em}}
\begin{document}

\title{Dimension Reduction for High Dimensional Vector Autoregressive Models}
\author{Gianluca Cubadda\thanks{Universita' di Roma "Tor Vergata", Dipartimento di
Economia e Finanza, Via Columbia 2, 00133 Roma, Italy. Email:
gianluca.cubadda@uniroma2.it.}\\Universit\`{a} di Roma "Tor Vergata"
\and Alain Hecq\thanks{Maastricht University, Department of Quantitative Economics,
P.O.Box 616, 6200 MD Maastricht, The Netherlands. Email:
a.hecq@maastrichtuniversity.nl.}\\Maastricht University}
\date{\today\footnote{Various versions of this paper were presented at the
31$^{\text{th}}$ (EC)$^{\text{2}}$ Conference on "High dimensional modeling in
time series", the 9$^{\text{th}}$ Italian Congress of Econometrics and
Empirical Economics, the International Association for Applied Econometrics
2021 Annual Conference, the workshop on "Dimensionality Reduction and
Inference in High-Dimensional Time Series" at Maastricht University, the 15th
International Conference on Computational and Financial Econometrics, the 3rd
Italian Workshop of Econometrics and Empirical Economics: \textquotedblleft
High-dimensional and Multivariate Econometrics: Theory and
Practice\textquotedblright, and at a seminar at Sichuan University. We thank
the participants, as well as Massimiliano Marcellino, Sung K. Ahn and two
anonymous referees, for helpful comments and suggestions. The usual
disclaimers apply.}}
\maketitle

\begin{abstract}
This paper aims to decompose a large dimensional vector autoregessive (VAR)
model into two components, the first one being generated by a small-scale VAR
and the second one being a white noise sequence. Hence, a reduced number of
common components generates the entire dynamics of the large system through a
VAR structure. This modelling, which we label as the dimension-reducible VAR,
extends the common feature approach to high dimensional systems, and it
differs from the dynamic factor model in which the idiosyncratic component can
also embed a dynamic pattern. We show the conditions under which this
decomposition exists. We provide statistical tools to detect its presence in
the data and to estimate the parameters of the underlying small-scale VAR
model. Based on our methodology, we propose a novel approach to identify the
shock that is responsible for most of the common variability at the business
cycle frequencies. We evaluate the practical value of the proposed methods by
simulations as well as by an empirical application to a large set of US
economic variables.

\bigskip


\newpage

\end{abstract}

\section{Introduction}

For decades, the vector autoregressive (VAR) model is \textit{de facto} a
standard tool for investigating multivariate time series data.\footnote{It is
usual and correct to quote Sims (1980) for his fundamental contribution to the
VAR literature. An earlier reference is Quenouille (1957).} In
macroeconometrics, VARs are routinely used for forecasting, for extracting
co-movements such as the presence of cointegration, to test for Granger
causality as well as to perform structural analyses. However, the number of
VAR parameters to be estimated increases quadratically with the number of
variables and linearly with the number of lags. This quickly compromises the
estimation results outside the case of small scale models. There is currently
a growing interest for jointly modeling many variables and consequently to
look at the feasibility to work with large dimensional VAR models. Indeed, the
increase of data availability in economics and finance\footnote{Both in terms
of the number of series that are easily available and in the use of different
sampling frequencies (e.g. mixed frequency VARs, see Goetz \textit{et al}.,
2016).} is associated with the common belief that using more information with
high dimensional econometric and statistical models will improve our
understanding of the macroeconomy as well as forecast accuracy (see Boivin and
Ng, 2006 for counter examples). Consequently, as the increase of the number of
time series jointly considered in VARs cannot be too large compared to a given
number of observations, different attempts have been proposed in the
literature to curb the curse of dimensionality problem. These methods can be
gathered in two categories: dimension reduction approaches on the one hand and
regularization techniques on the other hand. We include in the latter group
both Bayesian methods (surveyed in Karlsson, 2013, Koop, 2018), although
Bayesian techniques are also used to estimate large reduced-rank VARs (see
Carriero \textit{et al}., 2011), and the more recent booming contributions on
penalized estimation of sparse VARs (Wilms and Croux, 2016, Hsu \textit{et
al}., 2008, Nicholson \textit{et al}., 2018, Davis \textit{et al}., 2016,
Smeekes and Wijler, 2018, Kock and Callot, 2015, Hecq \textit{et al.}, 2021).
The former group of methods, to which our paper wishes to contribute, includes
reduced rank techniques (Reinsel, 1983, Ahn and Reinsel, 1988, Carriero
\textit{et al}., 2011, Cubadda and Hecq, 2011, Cubadda and Hecq, 2021,
Bernardini and Cubadda, 2015) and the huge literature on factor models
(surveyed in Stock and Watson, 2016, and Lippi, 2018, 2019).

Differently from those contributions on system dimension reductions, we
provide a framework where the whole dynamics of the system is due to an
underlying small scale VAR model. Following Lam \textit{et al}. (2011) and Lam
and Yao (2012), we first decompose the large multivariate time series $Y_{t}$
into two parts: a linear function of a small scale dynamic component $x_{t}$
and a static component $\varepsilon_{t}$ that, and this is the key point that
makes our approach different from the usual dynamic factor model, is
unpredictable from the past. The capital $Y_{t}$ stresses that we start from a
potentially high-dimensional time series process whereas the small $x_{t}$
stresses that a small number of factors are responsible for the entire
dynamics of the system. Then we provide the conditions under which such a
dynamic component $x_{t}$ is generated by a small scale VAR model.
Particularly, we show that it is required that the large VAR model of series
$Y_{t}$ is endowed with both the serial correlation common feature (Engle and
Kozicki, 1993) and an index structure (Reinsel, 1983) in order to ensure that
the dynamic component $x_{t}$ follows a VAR model. Hence, we provide a link
between the factor modeling for high-dimensional time series and the
reduced-rank VAR approach, which, to the best of our knowledge, was not noted
before. This bridge allows us to unravel common cyclical features and to
impose their presence in large VARs.

Obviously, the decomposition that we consider might not exist. Based on the
eigenanalysis proposed in Lam \textit{et al}. (2011) and Lam and Yao (2012),
we provide statistical tools to verify whether there exists in series $Y_{t}$
a dynamic component $x_{t}$ that is generated by such a small scale VAR model
and to estimate the associated parameters. If this is the case, the forecasts
of $x_{t}$ can be used to predict the future realizations of the large
dimensional system $Y_{t}$ and structural shocks may be recovered from the
reduced form errors of the dynamic component only.

Our contribution can be related to the early literature on the analysis of
linear transformations of vector autoregressive moving average (VARMA)
processes, see, \textit{inter alia}, Kohn (1982), and L\"{u}tkepohl (1984a,
1984b). These previous contributions established that non-singular linear
transformations of a VARMA process still have a VARMA structure, and if some
restrictions on the relationships among variables apply, the VARMA model for
the linear transformation can be parsimonious enough to be used in empirical
applications. However, at the representation theory level the aim of our
analysis is rather to find the conditions that allow to model a large
dimensional VAR process through a small scale VAR of the same order without
any information loss. To the best of our knowledge, such issue has not been
tackled so far. So doing we also provide the basis for further extending the
link between the final equation representation and VAR models with reduced
rank restrictions (Cubadda \textit{et al.}, 2009).

The rest of the paper is organized as follows. Section 2 presents the main
results on the model representation as well as the restrictions that our new
modelling entails. We observe that, so far, the reduced-rank VAR and
multivariate autoregressive index model have been considered separately in the
literature. We find out that the combination of the two models allows for the
important dimension reduction in large VARs that we seek. We show that
information criteria can be used to determine the dimension of a small VAR
within a large dimensional system and we provide estimators for its
parameters. Section 3 conducts an extensive Monte Carlo study to evaluate the
finite sample properties of the proposed tools. Section 4 provides an
empirical application a large set of macroeconomic and financial time series
to illustrate the practical value of our approach. Based on our methodology,
we propose a novel approach to identify the shock that is responsible for most
of the common volatility at the business cycle frequency band. Finally,
Section 5 concludes.

\section{Theory}

This section starts by presenting the derivation of the proposed modelling,
then it discusses the statistical inference in a large dimensional framework.

\subsection{Model representation}

Let us assume that the $n$-vector time series $Y_{t}$ is generated by the
following second-order stationary VAR($p$) model%
\begin{equation}
Y_{t}=\sum_{j=1}^{p}\Phi_{j}Y_{t-j}+{\small u}_{t}, \label{LVAR}%
\end{equation}
where $t=1,...,T$, $\Phi_{j}$ is an $n\times n$ matrix for $j=1,...,p$ with
$\Phi_{p}\neq0$ such that the roots of $\det\left(  I_{n}-\sum_{j=1}^{p}%
\Phi_{j}z^{j}\right)  $ lie outside the unit circle; ${\small u}_{t}$ is an
$n$-vector of errors such that $\mathrm{E}({\small u}_{t})=0$, $\mathrm{E}%
({\small u}_{t}{\small u}_{t}^{\prime})=\Sigma_{{\small u}}$ is a finite and
positive definite matrix, $\mathrm{E}({\small u}_{t}|\digamma_{t-1})=0$ and
$\digamma_{t}$ is the natural filtration of the process $Y_{t}$. For the sake
of simplicity, we assume that deterministic elements are absent (or that the
series have been demeaned or detrended first).

\begin{assumption}
For $\Phi=[\Phi_{1},...,\Phi_{p}]^{\prime}$\ it holds that $\Phi^{\prime}%
=\bar{A}\bar{\Omega}^{\prime},$ where $\bar{A}$ is a full rank $n\times r$
$(r<n$ $)$ matrix and $\bar{\Omega}=[\bar{\omega}_{1}^{\prime},...,\bar
{\omega}_{p}^{\prime}]^{\prime}$ is a full rank $np\times r$ matrix. Since we
can always use the equivalent factorization $\Phi^{\prime}=A\Omega^{\prime}$,
where $A=\bar{A}(\bar{A}^{\prime}\bar{A})^{-1/2}$ and $\Omega=\bar{\Omega
}(\bar{A}^{\prime}\bar{A})^{1/2}$, we assume without loss of generality that
$A$ is a matrix with orthogonal columns, namely $A^{\prime}A=I_{r}$.
\end{assumption}

Assumption 1 is popularly known in time series econometrics as the serial
correlation common feature (Engle and Kozicki, 1993). It was extensively
studied in connection with cointegration (see, \textit{inter alia}, Vahid and
Engle, 1993, Ahn, 1997, Cubadda and Hecq, 2001, Hecq \textit{et al}. 2006,
Cubadda, 2007, and Athanasopoulos \textit{et al.,} 2011). Moreover, it implies
that the marginal processes of series $Y_{t}$\ follow parsimonious univariate
models, thus solving the so-called autoregressivity paradox (Cubadda
\textit{et al}., 2009). See Centoni and Cubadda (2015) and Cubadda and Hecq
(2021) for recent surveys. In the analysis that follows, we focus on the case
where $n$ is large, virtually with a similar magnitude as the sample size $T$,
whereas $r$ is small compared to $T$.

We start by noting that under Assumption 1 we can use the identity
\begin{equation}
AA^{\prime}+A_{\bot}A_{\bot}^{\prime}=I_{n} \label{Ident}%
\end{equation}
in order to decompose series $Y_{t}$ as%
\begin{equation}
Y_{t}=Ax_{t}+\varepsilon_{t}, \label{decomp}%
\end{equation}
where $x_{t}=A^{\prime}Y_{t}$, $\varepsilon_{t}=A_{\bot}A_{\bot}^{\prime
}{\small u}_{t}$ and $A_{\bot}$ is a full-rank $n\times(n-r)$ matrix such that
$A_{\bot}^{\prime}A=0$ and $A_{\bot}^{\prime}A_{\bot}=I_{n-r}$.

Following Lam \textit{et al}. (2011) and Lam and Yao (2012), we call $x_{t}$
and $\varepsilon_{t}$ respectively\ the dynamic and the static component of
series $Y_{t}$. Indeed, we have for the disturbances $\varepsilon_{t}$%
\[
\mathrm{E}(\varepsilon_{t}|\digamma_{t-1})=A_{\bot}^{\prime}A_{\bot}%
\mathrm{E}({\small u}_{t}|\digamma_{t-1})=0
\]
from which it follows%
\begin{equation}
\mathrm{E}(Y_{t+k}|\digamma_{t})=A\mathrm{E}(x_{t+k}|\digamma_{t}),
\label{forecast}%
\end{equation}
where $k$, the forecast horizon, is any positive integer.

\begin{remark}
Note that the assumption that the matrix $\Omega$ as defined in Assumption 1
has full column rank is equivalent to require that no linear combinations of
$x_{t}$ are innovations w.r.t the past. Hence, decomposition $(\ref{decomp})$
allows us to disentangle the latent autocorrelated component $x_{t}$, whose
dimension cannot be further reduced, from the unpredictable component
$\varepsilon_{t}$.
\end{remark}

\begin{remark}
Representation $(\ref{decomp})$ has some analogies with a factor model but
there are substantial differences. First, the static component $\varepsilon
_{t}$ differs from idiosyncratic shocks in approximate factor models
(Chamberlain and Rothschild, 1983) in that the former is singular being driven
by the $(n-r)$ shocks $A_{\bot}^{\prime}{\small u}_{t}$ whereas the latter are
only mildly cross-correlated.\footnote{This difference has important
implications on the inferential properties of the two methodologies. Indeed,
Lam \textit{et al}. (2011) show by simulations that their approach outperforms
principal component methods when strong cross-correlations exist in the
noise.} Second, the dynamic component $x_{t}$ drives all the dynamics of the
system whereas the idiosyncratic components can also embed a dynamic pattern.
Third, in most of the factor literature it is assumed that the factors and the
idiosyncratic components are uncorrelated at any lead and lag whereas in
Equation $(\ref{decomp})$ we have that $\mathrm{E}(\varepsilon_{t+k}%
x_{t}^{\prime})=0$ only for $k>0$.
\end{remark}

Lam \textit{et al}. (2011) and Lam and Yao (2012) showed how to consistently
estimate both $r$ and $A$ (or, more formally, an $n\times r$\ matrix that lies
in the space generated by the columns of $A$) under assumptions that are
compatible with those in Assumption 1 even when the dimension $n$
diverges.\footnote{Notice that the aforementioned contributors assume that
$\mathrm{E}(\varepsilon_{t}x_{t}^{\prime})=0$. We tackle with this issue in
Remark 4.} Their method is simple to perform since it is based on the
eigenanalysis of the sum of the squared autocovariance matrices of series
$Y_{t}$. Having obtained an estimator, say $\hat{A}$, it is then possible to
get the dynamic and static components in (\ref{decomp}) respectively as
\[
\hat{x}_{t}=\hat{A}^{\prime}Y_{t}%
\]
and%
\[
\hat{\varepsilon}_{t}=(I-\hat{A}\hat{A}^{\prime})Y_{t}.
\]
\qquad

However, in order to forecast series $Y_{t},$ as well as to perform structural
analysis, estimating the loading matrix $A$ is not enough. In a dimension
reduction perspective, we further look at the conditions under which the
dynamic component $x_{t}$\ is generated by a small-scale VAR($p$) model. To
the best of our knowledge, such goal has not been pursued so far.

We start by noting that under Assumption 1 we can rewrite model (\ref{LVAR})
as follows%
\begin{equation}
Y_{t}=\sum_{j=1}^{p}A\omega_{j}^{\prime}Y_{t-j}+{\small u}_{t}, \label{RRLVAR}%
\end{equation}
which is popularly known as the reduced-rank VAR model (RRVAR) and it was
extensively analyzed, \textit{inter alia}, by Velu \textit{et al.} (1986) and
Ahn and Reinsel (1988). We have already emphasized that restrictions such as
(\ref{RRLVAR}) are at the heart of the serial correlation common feature literature.

Second, pre-multiplying both sides of Equation (\ref{RRLVAR}) by $A^{\prime}$
we get
\begin{equation}
x_{t}=\sum_{j=1}^{p}\omega_{j}^{\prime}Y_{t-j}+\xi_{t}, \label{VARX}%
\end{equation}
where $\xi_{t}=A^{\prime}{\small u}_{t}$. However, it is important to see that
Equation (\ref{VARX}) does not yet provide a small-scale model for series
$Y_{t}$. Indeed, $\Omega$ being a $np\times r$ matrix with elements $\omega$
in (\ref{VARX}), the number of parameters still grows proportionally with $n$,
although not with $n^{2}$ as in the unrestricted VAR.

Finally, we insert Equation (\ref{decomp}) into (\ref{VARX}) such that we
obtain%
\[
x_{t}=\sum_{j=1}^{p}\omega_{j}^{\prime}Ax_{t-j}+\sum_{j=1}^{p}\omega
_{j}^{\prime}\varepsilon_{t-j}+\xi_{t},
\]
which allows us to derive the condition under which the dynamic component
$x_{t}$ is generated by a VAR($p$) process as follows:

\begin{assumption}
For any $j=1,...,p$ it holds that $\omega_{j}\in\mathrm{Sp}(A)$, where
$\mathrm{Sp}(A)$\ indicates the space generated by the columns of $A$. Notice
that this is equivalent to require that $\omega_{j}=A\alpha_{j}^{\prime}$,
where $\alpha_{j}$ is a $r\times r$ matrix.
\end{assumption}

Indeed, under Assumption 2 we have that $\omega_{j}^{\prime}Ax_{t-j}%
=\alpha_{j}x_{t-j}$ and $\omega_{j}^{\prime}\varepsilon_{t-j}=0$ for
$j=1,...,p$, hence the data generating process of the dynamic component
$x_{t}$ boils down to%
\begin{equation}
x_{t}=\sum_{j=1}^{p}\alpha_{j}x_{t-j}+\xi_{t}. \label{var for x}%
\end{equation}
The intuition behind the algebra is that Assumption 2 requires that the lags
of the same linear combinations of $Y_{t}$ that are unpredictable from the
past are also irrelevant predictors of the dynamic component $x_{t}$.

Remarkably, the RRVAR model (\ref{RRLVAR}) of series $Y_{t}$\ can be rewritten
as
\begin{equation}
Y_{t}=\sum_{j=1}^{p}A\alpha_{j}\underset{x_{t-j}}{\underbrace{A^{\prime
}Y_{t-j}}}+{\small u}_{t}. \label{RR-MAI}%
\end{equation}
Model (\ref{RR-MAI}) is interesting since it combines the features of the
RRVAR model with those of the multivariate autoregressive index (MAI) model
proposed by Reinsel (1983). Recently, there has been a renewed interest in the
MAI because it allows to rewrite the VAR in a similar way as the popular
dynamic factor model, see \textit{inter alia} Carriero \textit{et al.} (2016),
Cubadda \textit{et al}. (2017), and Cubadda and Guardabascio (2019). So far,
the RRVAR and the MAI have been considered separately in the literature
whereas Assumption 2 reveals that the combination of the two models allows for
an important dimension reduction in large VARs. In what follows, we call model
(\ref{RR-MAI}) as the dimension-reducible VAR model (DRVAR).

\begin{remark}
Another popular approach in econometrics that aims at exploiting the
information of large dimensional time series is the factor augmented VAR
(FAVAR) as originally proposed by Bernanke et al. (2005). In such modelling,
first some unobserved factors are extracted from an high dimensional time
series, then it is assumed that these factors along with a small set of key
observed variables jointly follow a small-scale VAR model. Under the
assumption that the joint data generating process (DGP) of the observed
variables $Y_{t}$ is (\ref{LVAR}), it follows that the FAVAR is a restricted
case of the DRVAR with%
\begin{equation}
A=\left[
\begin{array}
[c]{cc}%
I_{m} & 0_{(n-m)\times m}\\
0_{m\times(r-m)} & B_{(n-m)\times(r-m)}%
\end{array}
\right]  \label{FAVAR}%
\end{equation}
such that $x_{t}=[y_{t}^{\prime},w_{t}^{\prime}]^{\prime}$, where
$y_{t}=[y_{1,t},...,y_{m,t}]^{\prime}$ is the $m$-vector ($m\leq r$) of the
key observed variables and $w_{t}=B^{\prime}[y_{m+1,t},...,y_{n,t}]^{\prime}$
is an $(r-m)$-vector of unobserved factors. Whether restrictions
$(\ref{FAVAR})$ are generally valid is an empirical issue. However, we remark
that a different choice of the key variables would induce a different set of
restrictions on the $A$ matrix. Since such choice is of course arbitrary, the
FAVAR reveals to have a rather ad hoc structure.
\end{remark}

In order to perform structural analysis through the DRVAR, one way to go is
inverting the polynomial VAR coefficient matrix in Equation (\ref{RR-MAI}) to
obtain the Wold representation of series $Y_{t}$. Here we offer an alternative
route. If we invert the polynomial coefficient matrix in Equation
(\ref{var for x}) and insert the Wold representation of the dynamic components
$x_{t}$\ in decomposition (\ref{decomp}) we get
\begin{equation}
Y_{t}=A\gamma(L)\xi_{t}+\varepsilon_{t}, \label{R_Wold}%
\end{equation}
where $\gamma(L)^{-1}=I_{n}-\sum_{j=1}^{p}\alpha_{j}L^{j}$. Finally, by
linearly projecting $\varepsilon_{t}$ on $\xi_{t}$, we can decompose the
static component as $\varepsilon_{t}=\rho\xi_{t}+\nu_{t}$, where $\rho
=A_{\bot}A_{\bot}^{\prime}\Sigma_{{\small u}}A(A^{\prime}\Sigma_{{\small u}%
}A)^{-1}$, and then rewrite Equation (\ref{R_Wold}) as%
\begin{equation}
Y_{t}=\underset{\chi_{t}}{\underbrace{C(L)\xi_{t}}}+\nu_{t}, \label{Sdecomp}%
\end{equation}
where $C_{0}=A+\rho$\ and $C_{j}=A\gamma_{j}$ for $j>0$.

Representation (\ref{Sdecomp}) highlights that the system dynamics are
entirely generated by errors $\xi_{t}$. Hence, we label $\chi_{t}$ as the
common component of $Y_{t}$ and $\nu_{t}$\ as the ignorable errors, as we
assume the latter are not endowed with a structural interpretation. Since the
errors $\xi_{t}$\ and $\nu_{t}$ are uncorrelated at any lead and lag, it is
then possible to recover the structural shocks solely from the reduced form
errors $\xi_{t}$ of the common component $\chi_{t}$ using any of the
procedures that are commonly employed in structural VAR analysis.

For instance, one may obtain the structural shocks as $u_{t}=C^{-1}D\xi_{t}$
and the impulse response functions from $\Psi(L)=C(L)D^{-1}C$, where $D$ is
the matrix formed by the first $r$ rows of $C_{0}$ and $C$ is a lower
triangular matrix such that $CC^{\prime}=DA^{\prime}\Sigma_{{\small u}%
}AD^{\prime}$. Since the first $r$ rows of $\Psi(0)$, being equal to $C$, form
a lower triangular matrix, the usual interpretation of the structural shocks
$u_{t}$ applies as long as the $s$ $(s\leq r)$ variables of interest are
placed and properly ordered in the first $s$ elements of $Y_{t}$. Notice that
such identification strategy is based on a unique rotation of the reduced form
common shocks $\xi_{t}$, and hence it does not require to endow the dynamic
component $x_{t}$ with an economic interpretation.

Clearly, it is always possible to identify the structural shocks directly from
the reduced form errors ${\small u}_{t}$ of the large VAR. However, the
advantage of the approach based on representation (\ref{Sdecomp}) is that it
requires to identify $r$ shocks only instead of $n$ of them. Hence, in the
structural DRVAR analysis based on (\ref{Sdecomp}) we have a number of
structural shocks that is smaller than the number of variables, as it is
typical in both structural factor models (see e.g. Forni \textit{et al}, 2009)
and dynamic stochastic general equilibrium (DSGE) models (see e.g.
Fern\'{a}ndez-Villaverde \textit{et al}., 2016).

\subsection{Statistical inference}

In order to consistently estimate the matrix $A$, we start by relying on the
approach suggested by Lam \textit{et al}. (2011), and Lam and Yao (2012). This
approach has recently been extended in various directions, such as
cointegration (Zhang \textit{et al}, 2019), principal component analysis for
stationary time series (Chang \textit{et al}, 2018), and multivariate
volatilities modelling (Tao \textit{et al}. 2011; Li \textit{et al}., 2016).

Let us denote the autocovariance matrix of series $Y_{t}$ at lag $j$ as
$\Sigma_{y}(j)=\mathrm{E}(Y_{t}Y_{t-j}^{\prime})$. In view of Equation
(\ref{decomp}), we see that%
\[
A_{\bot}^{\prime}\Sigma_{y}(j)=\mathrm{E}(A_{\bot}^{\prime}\varepsilon
_{t}Y_{t-j}^{\prime})=0\text{ \ \ }\forall j>0.
\]
Hence, the matrix $A$ lies in the space generated by the eigenvectors
associated with the $r$ non-zero eigenvalues of the symmetric and
semi-positive definite matrix%
\[
M=\sum_{j=1}^{p_{0}}\Sigma_{y}(j)\Sigma_{y}(j)^{\prime},
\]
where $p_{0}$ is a positive integer. Given the assumption that series $Y_{t}$
follow a finite order VAR($p$) model, one would ideally fix $p_{0}=p$.

Let us indicate with $\hat{V}_{q}$ the matrix formed by the eigenvectors
associated with the $q$ $(\leq n)$ largest eigenvalues of the matrix
\[
\hat{M}=\sum_{j=1}^{p_{0}}\hat{\Sigma}_{y}(j)\hat{\Sigma}_{y}(j)^{\prime},
\]
where $\hat{\Sigma}_{y}(j)$ denotes the sample autocovariance matrix of
$Y_{t}$ at lag $j$.

Under regularity conditions that are compatible with our assumptions, $\hat
{V}_{r}$\ estimates $A$\ (up to an orthonormal transformation) with a rate
equal to $n^{-\delta}T^{1/2}$ when $r$ is fixed, $n,T\rightarrow\infty$, and
$\bar{a}_{i}^{\prime}\bar{a}_{i}\asymp n^{1-\delta}$ for $i=1,...,r$, where
$\bar{A}=[\bar{a}_{1},...,\bar{a}_{r}]$ and $\delta\in\lbrack0,1]$. Notice
that $\delta$\ can be interpreted as an inverse measure of the strength of the
factors: when $\delta=0$\ the factors are strong since the common component is
shared by most of the $n$\ time series, whereas the factors are weak when
$\delta\in(0,1]$ (see Theorem 1 of Lam \textit{et al}., 2011). Moreover, Lam
and Yao (2012) proved that a consistent estimator of $r$ is provided by
\begin{equation}
\hat{r}=\arg\min_{i=1,..R}\left\{  \hat{\lambda}_{i+1}/\hat{\lambda}%
_{i}\right\}  , \label{r^hat}%
\end{equation}
where $R$ is a constant such that $r<R<n$ and $\hat{\lambda}_{i}$ is the
$i-$th largest eigenvalue of matrix $\hat{M}$.\footnote{Li \textit{et al}.
(2017) proposed an improved estimator of $r$ that is consistent even when not
all the factors have the same strength. However, they take the assumption of
independence between $x_{t}$ and $\varepsilon_{t}$, which clearly does not
hold here.}

\begin{remark}
Lam \textit{et al}. (2011) and Lam and Yao (2012) assume that $\mathrm{E}%
(\varepsilon_{t+k}x_{t}^{\prime})=0$ for $k\geq0$, whereas in our framework
the strict inequality $k>0$\ only holds. However, we can always transform the
original decomposition $(\ref{decomp})$ in such a way that the two components
are contemporaneously uncorrelated and the static component is still a white
noise. Indeed, using the decomposition of the identity matrix proposed by
Centoni and Cubadda (2003)%
\[
A\underset{\breve{A}^{\prime}}{\underbrace{(A^{\prime}\Sigma_{{\small u}}%
^{-1}A)^{-1}A^{\prime}\Sigma_{{\small u}}^{-1}}}+\underset{\breve{A}_{\bot
}}{\underbrace{\Sigma_{{\small u}}A_{\perp}(A_{\perp}^{\prime}\Sigma
_{{\small u}}A_{\perp})^{-1}}}A_{\perp}^{\prime}=I_{n}%
\]
we can decompose series $Y_{t}$\ as%
\[
Y_{t}=A\breve{x}_{t}+\breve{\varepsilon}_{t},
\]
where $\breve{x}_{t}=\breve{A}^{\prime}Y_{t}$, $\breve{\varepsilon}_{t}%
=\breve{A}_{\bot}A_{\bot}^{\prime}{\small u}_{t}$. It follows that%
\[
\mathrm{E}(\breve{\varepsilon}_{t}\breve{x}_{t}^{\prime})=\breve{A}_{\bot
}A_{\bot}^{\prime}\Sigma_{{\small u}}\breve{A}=0
\]
and%
\[
\mathrm{E}(\breve{\varepsilon}_{t}|\digamma_{t-1})=\breve{A}_{\bot}A_{\bot
}^{\prime}\mathrm{E}({\small u}_{t}|\digamma_{t-1})=0.
\]
Hence, in our framework the assumption that the dynamic component and the
white noise are contemporaneously uncorrelated turns out to be unnecessary for
the estimation of $A$ and $r$ through the eigen-analysis of the matrix $M$.
\end{remark}

\begin{remark}
When the factors are strong, i.e. $\delta=0$, and the cross-correlation
between the dynamic and static component is not so large to distort the
information on the autocorrelation of the former, a "blessing of
dimensionality" phenomenon occurs since the estimating accuracy of $\hat
{V}_{r}$ has the standard $\sqrt{T}$ rate independently from the dimension
$n$. The intuition is that the strong factors exploit the information coming
from most of, if not all, the $n$ series, hence the curse of dimensionality is
offset by the increase of the information on the dynamic component (for
further details and comments see Section 3 of Lam \textit{et al}., 2011).
\end{remark}

Notice that $\hat{r}$ in (\ref{r^hat}) consistently estimates the rank of the
matrix $M$ when Assumption 1 only applies, whereas we need an estimator of $r$
that is subject to Assumption 2 as well. Let us first consider the problem of
estimating the parameters of model (\ref{RR-MAI}) assuming that $r$ is known
and having fixed $A$ equal to $\hat{V}_{r}$. In order to accomplish this goal,
it is convenient to rewrite model (\ref{RR-MAI}) in its matrix form%
\begin{equation}
Y=Z\alpha A^{\prime}+{\small u}, \label{Y}%
\end{equation}
where $Y=\left[  y_{p+1},...,y_{T}\right]  ^{\prime}$, ${\small u}=\left[
{\small u}_{p+1},...,{\small u}_{T}\right]  ^{\prime}$, $z_{t}=\left[
x_{t}^{\prime},...,x_{t-p+1}^{\prime}\right]  ^{\prime}$, and $Z=\left[
z_{p},...,z_{T-1}\right]  ^{\prime}$. Then apply the $\mathrm{Vec}$ operator
to both the sides of Equation (\ref{Y}) and use the property $\mathrm{Vec}%
(ABC)=(C^{\prime}\otimes A)\mathrm{Vec}(B)$ to get%
\begin{equation}
\mathrm{Vec}(Y)=\left(  A\otimes Z\right)  \mathrm{Vec}(\alpha)+\mathrm{Vec}%
({\small u}), \label{Vec(Y)}%
\end{equation}
from which it is easy to see that the ordinary least squares (OLS) estimator
of $\mathrm{Vec}(\alpha)$\ in Equation (\ref{Vec(Y)}) takes the following
form:%
\begin{equation}
\mathrm{Vec}(\hat{\alpha})=[A^{\prime}\otimes(Z^{\prime}Z)^{-1}Z^{\prime
}]\mathrm{Vec}(Y). \label{Vec(alpha_hat)}%
\end{equation}
The main theoretical justification for considering the estimator
(\ref{Vec(alpha_hat)}) is that it is equivalent to applying OLS on
(\ref{var for x}),\footnote{This result is obtained by post-multiplying with
$A$ both sides of Equation (\ref{Y}) and then applying the Vec operator to get%
\[
\mathrm{Vec}(YA)=(I_{r}\otimes Z)\mathrm{Vec}(\alpha)+\mathrm{Vec}(\epsilon
A)
\]
It is easy to see that the OLS estimator of $\mathrm{Vec}(\alpha)$ in the
model above is the same as (\ref{Vec(alpha_hat)}).} which turns out to be the
quasi maximum likelihood (QML) estimator of parameters $\alpha$ in the
small-scale VAR model of the factor $x_{t}$ under the assumption that $A$\ is known.

An alternative estimator of parameters $\alpha$ can be obtained by applying
the generalized least squares (GLS) on Equation (\ref{Vec(Y)}). In particular,
pre-multiply both the sides of Equation (\ref{Vec(Y)}) by $\Sigma_{{\small u}%
}^{-1/2}\otimes I_{T-p}$ to get%
\begin{equation}
(\Sigma_{{\small u}}^{-1/2}\otimes I_{T-p})\mathrm{Vec}(Y)=\left(
\Sigma_{{\small u}}^{-1/2}A\otimes Z\right)  \mathrm{Vec}(\alpha
)+(\Sigma_{{\small u}}^{-1/2}\otimes I_{T-p})\mathrm{Vec}({\small u}).
\label{Vec(Y*Sigma^-.5)}%
\end{equation}
Tedious but simple algebra reveals that the OLS estimator of $\mathrm{Vec}%
(\alpha)$\ in Equation (\ref{Vec(Y*Sigma^-.5)}) takes the following form:%
\begin{equation}
\mathrm{Vec}(\tilde{\alpha})=\left[  \left(  A^{\prime}\Sigma_{{\small u}%
}^{-1}A\right)  ^{-1}A^{\prime}\Sigma_{{\small u}}^{-1}\otimes\left(
Z^{\prime}Z\right)  ^{-1}Z^{\prime}\right]  \mathrm{Vec}(Y).
\label{Vec(alpha_tilde)}%
\end{equation}
In view of Equation (\ref{Vec(Y*Sigma^-.5)}), it is easy to see that the GLS
estimator (\ref{Vec(alpha_tilde)}) is the QML estimator of parameters $\alpha$
in model (\ref{Y}) under the assumption that $A$ and $\Sigma_{{\small u}}$ are known.

The relation, in terms of efficiency, between the estimators
(\ref{Vec(alpha_hat)}) and (\ref{Vec(alpha_tilde)}) is provided in the
following theorem.

\begin{theorem}
Assuming that $A$ and $\Sigma_{{\small u}}$ are known, estimator
(\ref{Vec(alpha_tilde)}) of $\mathrm{Vec}(\alpha)$ has a mean square error
matrix, conditionally on $Z$, that is not larger than the one of estimator
(\ref{Vec(alpha_hat)}). The two estimators have the same mean square error
matrix when $A^{\prime}{\small u}_{t}$ and $A_{\bot}^{\prime}{\small u}_{t}$
are not correlated.
\end{theorem}

\begin{proof}
See the appendix.
\end{proof}

In order to derive a feasible GLS (FGLS) estimator, we suggest the following
switching algorithm, which has the property to increase the Gaussian
likelihood conditional to $A$ in each step.

\begin{enumerate}
\item In view of Equation (\ref{RR-MAI}) and given (initial) estimates of
$\alpha$, maximize the conditional Gaussian likelihood $%
%TCIMACRO{\tciLaplace}%
%BeginExpansion
\mathcal{L}%
%EndExpansion
(\Sigma_{{\small u}}|\alpha,A)$ by estimating $\Sigma_{{\small u}}$ with
\[
\left(  T-p\right)  ^{-1}(Y^{\prime}-A\alpha^{\prime}Z^{\prime})(Y-Z\alpha
A^{\prime}).
\]

\item Given the previously obtained estimate of $\Sigma_{{\small u}}$,
maximize $%
%TCIMACRO{\tciLaplace}%
%BeginExpansion
\mathcal{L}%
%EndExpansion
(\alpha|\Sigma_{{\small u}},A)$ by estimating elements of $\alpha$ with
(\ref{Vec(alpha_tilde)}).

\item Repeat steps 1 and 2 till numerical convergence occurs.\footnote{A
general proof of the convergence of this family of iterative procedures is
given by Oberhofer and Kmenta (1974).}
\end{enumerate}

In order to speed up the numerical convergence of that algorithm, it is
important to choose the initial values for the coefficient matrix $\alpha$
correctly. An obvious choice is resorting to $\hat{\alpha}$, which provides a
consistent estimate of $\alpha$ as $T$ increases.

A practical problem that arises when the sample size $T$ and the dimension $n$
are of similar magnitude is that the estimate of matrix $\Sigma_{{\small u}}%
$\ is singular or nearly singular. We propose to solve this problem by
ignoring the error cross-correlations in the estimation method. In particular,
we suggest to use a diagonal matrix $\Delta_{{\small u}}$ with the same
diagonal as $\Sigma_{{\small u}}$ in place of $\Sigma_{{\small u}}$ itself in
the FGLS procedure. This solution has two main motivations. First, it makes
the objective function of the switching algorithm to become $\mathrm{trace}%
(\ln(\Delta_{{\small u}}))$, which is a common approximation of $\ln
(\det(\Sigma_{{\small u}}))$ in high-dimensional settings, see Hu \textit{et
al}. (2017) and the references therein. Second, it is reasonable to presume
that the fraction of unanticipated co-movements among variables is small when
the conditioning information set is large.

Finally, in order to identify the dimension the dynamic component $r,$ we
suggest the following strategy. For $q=1,...,R$ estimate either by OLS or FGLS
the models%
\[
Y_{t}=\sum_{j=1}^{p}\hat{V}_{q}\alpha_{j,q}\hat{V}_{q}^{\prime}Y_{t-j}%
+{\small u}_{t}(q),
\]
where $\alpha_{j,q}$ is a $q\times q$ matrix for $j=1,...,p$, and estimate $r$
as the index $\widehat{r}$ that minimizes an information criterion such as
\begin{equation}
\mathrm{IC}(q)=n^{-1}\ln%
%TCIMACRO{\tprod \limits_{i=1}^{n}}%
%BeginExpansion
{\textstyle\prod\limits_{i=1}^{n}}
%EndExpansion
\hat{\sigma}_{i}^{2}(q)+\frac{c_{T}k}{Tn} \label{IC}%
\end{equation}
where $\hat{\sigma}_{i}^{2}(q)=(T-p)^{-1}%
%TCIMACRO{\tsum \limits_{t=p+1}^{T}}%
%BeginExpansion
{\textstyle\sum\limits_{t=p+1}^{T}}
%EndExpansion
{\small u}_{t,i}^{2}(q)$, ${\small u}_{t}(q)=[{\small u}_{t,1}%
(q),...,{\small u}_{t,n}(q)]^{\prime}$, $k=nq+(p-1)q^{2}\mathtt{,}$ $c_{T}%
$\ is a penalty term such that $c_{T}=2$ for the Akaike information criterion
(AIC), $c_{T}=2\ln(\ln(T))$ for the Hannan-Quinn information criterion (HQIC),
and $c_{T}=\ln(T)$ for the Bayes information criterion (BIC). Notice that the
measure of fit to be used is $\mathrm{trace}(\ln(\Delta_{{\small u}}))$, given
the assumption that $\Sigma_{{\small u}}=\Delta_{{\small u}}$, and the overall
number of parameters is $k=nq+(p-1)q^{2}$, given that the number of free
parameters in a base of the space spanned by $A$ is equal to $nq-q^{2}$ and
each of the $p$ $\alpha_{j,q}$ matrices has $q^{2}$ coefficients$.$

The asymptotic behavior of the information criteria (\ref{IC}) is given in the
following proposition.

\begin{proposition}
Under conditions such that OLS and FGLS estimate\ the DRVAR parameters (up to
an orthonormal transformation) with the standard $\sqrt{T}$ rate as
$n,T\rightarrow\infty$, and assuming that $\gamma=\underset{n\rightarrow
\infty}{\lim}\left(
%TCIMACRO{\tprod \limits_{i=1}^{n}}%
%BeginExpansion
{\textstyle\prod\limits_{i=1}^{n}}
%EndExpansion
\sigma_{i}^{2}\right)  ^{1/n}$ exists, where $\mathrm{diag}[\sigma_{1}%
^{2},...,\sigma_{n}^{2}]=\Delta_{{\small u}}$, the BIC and HQIC provide weakly
consistent estimators for the number of dynamic components $r$ but not for the
overall number of the DRVAR parameters $k$.
\end{proposition}

\begin{proof}
See the appendix.
\end{proof}

\section{Monte Carlo analysis}

\subsection{The data generating process}

In this section we perform a Monte Carlo study to evaluate the finite sample
performances of the OLS and FGLS estimators of model (\ref{RRLVAR}) parameters
having estimated the matrix $A$ according to Lam \textit{et al.} (2011) in
both cases. We consider the following $n$-dimensional stationary VAR$(2)$
process
\begin{equation}
Y_{t}=\underset{\Phi_{1}}{\underbrace{\bar{A}\mathrm{diag}(\delta_{1})\bar
{A}^{+}}}Y_{t-1}+\underset{\Phi_{2}}{\underbrace{\bar{A}\mathrm{diag}%
(\delta_{2})\bar{A}^{+}}}Y_{t-2}+{\small u}_{t}, \label{LVAR(2)}%
\end{equation}
where $\bar{A}$ is a $n\times r$ matrix such that its columns are generated by
$r$ i.i.d. $\mathrm{N}_{n}(0,I_{n})$, $\bar{A}^{+}=(\bar{A}^{\prime}\bar
{A})^{-1}\bar{A}^{\prime}$ is the Moore--Penrose pseudo inverse of the matrix
$\bar{A}$, $\delta_{1}=2\mathrm{diag}(m)\cos(\omega)$, $m$\ is a $r-$vector
drawn from a $\mathrm{U}_{r}[0.3,0.9]$, $\omega$\ is a $r-$vector drawn from a
$\mathrm{U}_{r}[0,\pi]$, $\delta_{2}=-m^{2}$, and ${\small u}_{t}$ are i.i.d.
$\mathrm{N}_{n}(0,\Sigma_{{\small u}})$.\footnote{Notice that the squared
Euclidean norms of the columns of $\bar{A}$ are proportional to $n$ on average
over the replications. However, the randomness of the matrix $\bar{A}$ ensures
for each replication some variability of the degree of strength over the $r$
factors.} Notice that Equation (\ref{LVAR(2)}) is equivalent to%
\[
Y_{t}=A\alpha_{1}A^{\prime}Y_{t-1}+A\alpha_{2}A^{\prime}Y_{t-2}+{\small u}%
_{t},
\]
where $A=\bar{A}S^{-1}$, $S=(\bar{A}^{\prime}\bar{A})^{1/2}$, and $\alpha
_{i}=S\mathrm{diag}(\delta_{i})S^{-1}$ for $i=1,2$ and consequently imposes
Assumptions 1 and 2 of our specification.

In order to simulate series $Y_{t}$, we first generate the diagonal VAR$(2)$
process%
\[
\bar{x}_{t}=\mathrm{diag}(\delta_{1})\bar{x}_{t-1}+\mathrm{diag}(\delta
_{2})\bar{x}_{t-2}+\bar{A}^{+}{\small u}_{t}%
\]
and then the static component $\varepsilon_{t}=\bar{A}_{\bot}\bar{A}_{\bot
}^{+}{\small u}_{t}$ with $\eta_{t}=[(\bar{A}^{+}{\small u}_{t})^{^{\prime}%
},(\bar{A}_{\bot}^{+}{\small u}_{t})^{\prime}]^{\prime}$ that are i.i.d.
$\mathrm{N}_{n}(0,\Sigma_{\eta})$. We can finally obtain%
\begin{equation}
Y_{t}=\bar{A}\bar{x}_{t}+\varepsilon_{t}=Ax_{t}+\varepsilon_{t}, \label{DGP}%
\end{equation}
where $x_{t}=S\bar{x}_{t}$ is the dynamic component.

An important role in the data generating process is played by the covariance
matrix $\Sigma_{\eta}$, which has the following Toeplitz structure%
\[
\Sigma_{\eta}=\left[
\begin{array}
[c]{cccccc}%
1 & \tau & \tau^{2} & \tau^{3} & \cdots & \tau^{n}\\
\tau & 1 & \tau & \tau^{2} & \cdots & \tau^{n-1}\\
\vdots & \vdots & \vdots & \vdots & \vdots & \vdots\\
\tau^{n} & \tau^{n-1} & \tau^{n-2} & \tau^{n-3} & \cdots & 1
\end{array}
\right]  ,
\]
where $\tau$ is a scalar drawn from a $\mathrm{U}[-0.5,0.5]$. Notice that
since $\Sigma_{\eta}$ is not diagonal by implication the covariance matrix of
the VAR errors $\Sigma_{{\small u}}$ is not diagonal as well. This allows us
to evaluate the performances of both the FGLS estimator and the information
criteria when $\ln(\det(\Sigma_{{\small u}}))$ is not equal to $\mathrm{trace}%
(\ln(\Delta_{{\small u}}))$.

\subsection{Results}

From (\ref{DGP}) we generate systems of successively$\ n=150,300,600,1200$
variables. We consider $r=3,9$ and $n$ dynamic components; $r=3$ is indeed
often assumed in financial applications (the Fama-French factors) whereas
several studies find that there exist from around 8 up to 10 factors in large
macroeconomic datasets. The case $r=n$ is considered to evaluate the
performances of the various estimators of $r$ when Assumptions 1 and 2 are not
valid.\footnote{Notice that when $r=n$ the DGP(\ref{DGP}) boils down to
$Y_{t}=\bar{A}\bar{x}_{t}$, where $\bar{A}$ is a $n\times n$ matrix such that
its columns are generated by $n$ i.i.d. $\mathrm{N}_{n}(0,I_{n})$ and $\bar
{x}_{t}$ is generated by a $n-$dimensional diagonal VAR(2) process.}

The number of observations is successively $T=\frac{1}{2}n,n,1.5n$. We
consequently evaluate the performance of our approach when the number of
variables is respectively less, equal or larger than the sample size. We
simulate $T+50$ observations and the first $50$ points are used as a burn-in
period, the remaining ones for estimations. Results are based on 1000 replications.

The proposed methods are evaluated by means of two statistics. We first
compute the percentage with which the number of dynamic components $r$ is
either correctly identified ($\%\widehat{r}=r$) when $r=3,9$ or hits the upper
bound ($\%\widehat{r}=R=11$) when $r=n$ using both the estimator (\ref{r^hat})
proposed by Lam and Yao (2012), hereafter denoted as LY, and the usual
information criteria under the assumption that $\Sigma_{{\small u}}$ is a
diagonal matrix. We also compute the average of $\hat{r}$ over the
replications as well as the frequencies with which those procedures
underestimate the correct number of dynamic components ($\%\widehat{r}<r$)
when $r<n$. Second, when the DRVAR restrictions are valid, i.e. $r=3,9$, we
compute the Frobenius distance between the estimates of $\Phi=[\Phi_{1}%
,\Phi_{2}]^{\prime}$\ and the true ones relative to the Frobenius norm of
$\Phi$ (RFD) as a measure of estimation precision. We only document the
results for the LY procedure and the OLS estimator since those with FGLS are
rather similar to the OLS ones and are then omitted for the sake of
space.\footnote{In particular, the percentages of correct identification that
are obtained by each criterion are almost identical irrespective of the
estimation method, whereas FGLS generally exhibits slightly lower RFDs than
OLS, although the differences are significant in about one third of the cases.
Results are available upon request.} \begin{table}[ptb]
\caption{Monte Carlo results, $r=3$, OLS estimator}%
\label{MC_table_1}
\begin{center}%
\begin{tabular}
[c]{llcccccccc}\hline\hline
$T/n$ &  & \multicolumn{4}{c}{$n=150$} & \multicolumn{4}{c}{$n=300$%
}\\\cline{3-10}
&  & $\%\widehat{r}=3$ & $\%\widehat{r}<3$ & $\overline{\widehat{r}}$ & RFD &
$\%\widehat{r}=3$ & $\%\widehat{r}<3$ & $\overline{\widehat{r}}$ & RFD\\\hline
$T=\frac{n}{2}$ & LY & 40.8 & 57.1 & 2.100 & - & 53.7 & 43.0 & 2.389 & -\\
& BIC & 67.2 & 32.0 & 2.605 & 39.46 & 78.6 & 21.3 & 2.748 & 27.55\\
& HQIC & 72.6 & 12.1 & 3.094 & 74.88 & 86.6 & 7.5 & 2.993 & 42.44\\
& AIC & 29.1 & 3.2 & 5.983 & 274.27 & 42.5 & 1.7 & 4.801 & 192.21\\\hline
$T=n$ & LY & 52.8 & 45.1 & 2.311 & - & 62.3 & 35.2 & 2.493 & -\\
& BIC & 80.3 & 19.5 & 2.784 & 26.94 & 90.0 & 9.90 & 2.898 & 18.56\\
& HQIC & 86.2 & 7.6 & 3.002 & 38.11 & 92.7 & 4.1 & 2.992 & 23.72\\
& AIC & 45.1 & 1.4 & 4.633 & 144.22 & 51.1 & 0.7 & 4.444 & 122.71\\\hline
$T=1.5n$ & LY & 55.8 & 43.1 & 2.328 & - & 64.3 & 32.7 & 2.555 & -\\
& BIC & 83.6 & 16.2 & 2.824 & 22.15 & 91.8 & 8.2 & 2.917 & 14.916\\
& HQIC & 91.6 & 4.4 & 3.005 & 27.93 & 95.0 & 2.4 & 3.003 & 19.140\\
& AIC & 53.5 & 0.7 & 4.245 & 105.08 & 52.8 & 0.1 & 4.217 &
99.841\\\hline\hline
&  & \multicolumn{4}{c}{$n=600$} & \multicolumn{4}{c}{$n=1200$}\\
& IC & $\%\widehat{r}=3$ & $\%\widehat{r}<3$ & $\overline{\widehat{r}}$ &
RFD & $\%\widehat{r}=3$ & $\%\widehat{r}<3$ & $\overline{\widehat{r}}$ &
RFD\\\hline
$T=\frac{n}{2}$ & LY & 59.9 & 35.1 & 2.560 & - & 65.7 & 27.8 & 2.734 & -\\
& BIC & 89.6 & 10.4 & 2.882 & 18.39 & 94.0 & 6.00 & 2.936 & 12.89\\
& HQIC & 92.5 & 2.9 & 3.017 & 29.07 & 95.5 & 2.2 & 3.003 & 17.84\\
& AIC & 51.5 & 0.6 & 4.286 & 147.14 & 52.8 & 0.1 & 4.171 & 137.20\\\hline
$T=n$ & LY & 67.7 & 27.9 & 2.683 & - & 73.2 & 21.5 & 2.815 & -\\
& BIC & 94.0 & 5.8 & 2.944 & 13.06 & 97.6 & 2.4 & 2.976 & 8.85\\
& HQIC & 96.8 & 1.3 & 3.009 & 16.31 & 97.4 & 0.8 & 3.010 & 12.36\\
& AIC & 52.2 & 0.1 & 4.167 & 110.85 & 52.0 & 0.0 & 4.176 & 112.70\\\hline
$T=1.5n$ & LY & 73.5 & 22.6 & 2.755 & - & 77.5 & 18.2 & 2.819 & -\\
& BIC & 96.5 & 3.5 & 2.964 & 10.54 & 99.1 & 0.9 & 2.991 & 7.27\\
& HQIC & 96.4 & 0.8 & 3.021 & 15.08 & 97.7 & 0.0 & 3.025 & 11.72\\
& AIC & 50.8 & 0.0 & 4.162 & 98.16 & 52.3 & 0.0 & 4.127 & 94.88\\\hline
\multicolumn{10}{l}{Notes: Percentages with which each method correctly
estimates or underestimates the}\\
\multicolumn{10}{l}{true $r$, the average of \ estimates of $r$ over $1000$
replications, and the Frobenius distance}\\
\multicolumn{10}{l}{between $\Phi$ and its estimates relative to the Frobenius
norm of $\Phi$.}\\\hline\hline
\end{tabular}
\end{center}
\end{table}

\begin{center}
\begin{table}[ptb]
\caption{Monte Carlo results, $r=9$, OLS estimator}%
\label{MC_table_2}
\begin{center}%
\begin{tabular}
[c]{llcccccccc}\hline\hline
$T/n$ &  & \multicolumn{4}{c}{$n=150$} & \multicolumn{4}{c}{$n=300$%
}\\\cline{3-10}
&  & $\%\widehat{r}=9$ & $\%\widehat{r}<9$ & $\overline{\widehat{r}}$ & RFD &
$\%\widehat{r}=9$ & $\%\widehat{r}<9$ & $\overline{\widehat{r}}$ & RFD\\\hline
$T=\frac{n}{2}$ & LY & 22.7 & 77.3 & 3.639 & - & 39.2 & 60.8 & 4.753 & -\\
& BIC & 44.7 & 54.5 & 7.781 & 59.54 & 50.0 & 50.0 & 8.135 & 38.57\\
& HQIC & 63.1 & 9.8 & 9.262 & 109.20 & 85.7 & 9.1 & 8.946 & 48.29\\
& AIC & 18.7 & 0.3 & 10.405 & 202.09 & 35.2 & 0.1 & 10.059 & 153.15\\\hline
$T=n$ & LY & 34.5 & 65.5 & 4.443 & - & 47.1 & 52.9 & 5.389 & -\\
& BIC & 47.9 & 52.1 & 8.108 & 39.41 & 61.4 & 38.6 & 8.459 & 27.38\\
& HQ & 83.0 & 11.6 & 8.928 & 46.82 & 91.4 & 6.3 & 8.958 & 30.55\\
& AIC & 41.3 & 0.6 & 9.911 & 115.60 & 45.5 & 0 & 9.836 & 100.99\\\hline
$T=1.5n$ & LY & 42.7 & 57.3 & 5.075 & - & 56.8 & 43.2 & 6.019 & -\\
& BIC & 59.8 & 40.2 & 8.386 & 32.17 & 74.3 & 25.7 & 8.681 & 22.04\\
& HQIC & 90.2 & 7.6 & 8.940 & 34.47 & 94.8 & 3.9 & 8.974 & 23.45\\
& AIC & 48.9 & 0.2 & 9.765 & 88.17 & 51.4 & 0.0 & 9.729 & 78.47\\\hline\hline
&  & \multicolumn{4}{c}{$n=600$} & \multicolumn{4}{c}{$n=1200$}\\
& IC & $\%\widehat{r}=9$ & $\%\widehat{r}<9$ & $\overline{\widehat{r}}$ &
RFD & $\%\widehat{r}=9$ & $\%\widehat{r}<9$ & $\overline{\widehat{r}}$ &
RFD\\\hline
$T=\frac{n}{2}$ & LY & 51.1 & 48.9 & 5.534 & - & 59.5 & 40.5 & 6.162 & -\\
& BIC & 61.1 & 38.8 & 8.442 & 27.03 & 73.3 & 26.7 & 8.676 & 18.96\\
& HQIC & 91.7 & 6.2 & 8.953 & 30.69 & 96.1 & 3.4 & 8.970 & 19.98\\
& AIC & 41.6 & 0.2 & 9.893 & 131.91 & 46.9 & 0.0 & 9.800 & 119.15\\\hline
$T=n$ & LY & 64.4 & 35.6 & 6.540 & - & 74.2 & 25.8 & 7.164 & -\\
& BIC & 76.2 & 23.8 & 8.719 & 18.83 & 88.9 & 11.1 & 8.873 & 13.16\\
& HQIC & 96.8 & 2.4 & 8.984 & 19.85 & 99.5 & 0.3 & 8.998 & 13.47\\
& AIC & 48.8 & 0.0 & 9.760 & 90.40 & 45.6 & 0.0 & 9.818 & 93.59\\\hline
$T=1.5n$ & LY & 64.9 & 35.1 & 6.559 & - & 75.7 & 24.3 & 7.285 & -\\
& BIC & 83.1 & 16.9 & 8.809 & 15.25 & 93.1 & 6.9 & 8.927 & 10.58\\
& HQIC & 97.9 & 1.7 & 8.986 & 15.54 & 99.5 & 0.2 & 9.001 & 10.97\\
& AIC & 51.1 & 0.0 & 9.746 & 74.52 & 48.0 & 0.0 & 9.800 & 76.15\\\hline
\multicolumn{10}{l}{See the notes of Table \ref{MC_table_1}.}\\\hline\hline
\end{tabular}
\end{center}
\end{table}
\end{center}

We first examine the results for $r=3,9$, which are respectively reported in
Tables \ref{MC_table_1} and \ref{MC_table_2}. As expected, we notice that all
methods perform better as the dimension of the system $n$ increases and,
conditional to a given $n$, as the sample size $T$ gets larger. For what
regards the estimation of the true number of dynamic components, we see that
the information criteria outperform the LY procedure by a clear margin. This
finding is hardly surprising given the parametric nature of the information
criteria. In particular, HQIC identifies the correct model better than the
competitors but in 3 cases where BIC performs best. In contrast, LY [AIC]
systematically underestimates [overestimates] the true $r$. With respect to
the RFD, we observe that the models identified by the BIC [AIC] provide
estimates of $\Phi$ that are more [less] accurate than those obtained by the
other criteria over all the settings. Such outcomes are likely due the fact
that the BIC is downward biased, thus implicitly shrinking to zero the dynamic
components that are only mildly autocorrelated, whereas the AIC systematically
overfits the model, thus inflating the estimation error. This finding
discourages the use of the AIC in empirical applications. Overall, BIC and
HQIC tend to perform similarly in both identification and estimation precision
as both $n$ and $T$ get large.

Table \ref{MC_table_3} reports the results when the DRVAR restrictions are not
valid, i.e. for $r=n$. We see that the LY procedure, which is designed to
estimate $r$ when Assumption 1 is valid, spuriously suggests the presence of
about two dynamic components on average irrespective of the sample size and
the system dimension. In contrast, the usual information criteria, especially
HQIC and AIC, provide estimates of $r$ that get close or hit the upper bound
starting from $n,T=300,300$ up to larger values of both $n$ and $T$. These
findings indicate that, when there exist no common dynamic components in the
data, the information criteria (\ref{IC}) correctly provide estimates of $r$
that tend to become larger as $n,T$ increase.

\begin{table}[ptb]
\caption{Monte Carlo results, $r=n$, OLS estimator}%
\label{MC_table_3}
\begin{center}%
\begin{tabular}
[c]{llcccccccc}\hline\hline
$T/n$ &  & \multicolumn{2}{c}{$n=150$} & \multicolumn{2}{c}{$n=300$} &
\multicolumn{2}{c}{$n=600$} & \multicolumn{2}{c}{$n=1200$}\\\cline{3-10}
&  & $\%\widehat{r}=R$ & $\overline{\widehat{r}}$ & $\%\widehat{r}=R$ &
$\overline{\widehat{r}}$ & $\%\widehat{r}=R$ & $\overline{\widehat{r}}$ &
$\%\widehat{r}=R$ & $\overline{\widehat{r}}$\\\hline
$T=\frac{n}{2}$ & LY & 0.0 & 2.022 & 0.10 & 2.028 & 0.10 & 2.156 & 0.30 &
2.316\\
& BIC & 0.0 & 5.178 & 15.30 & 8.116 & 84.40 & 10.718 & 100.0 & 11.000\\
& HQIC & 46.4 & 9.638 & 94.60 & 10.914 & 100.0 & 11.000 & 100.0 & 11.000\\
& AIC & 96.7 & 10.962 & 99.80 & 10.998 & 100.0 & 11.000 & 100.0 &
11.000\\\hline
$T=n$ & LY & 0.30 & 2.032 & 0.10 & 2.252 & 0.40 & 2.281 & 0.50 & 2.443\\
& BIC & 13.00 & 7.939 & 85.40 & 10.744 & 100.0 & 11.000 & 100.0 & 11.000\\
& HQIC & 85.4 & 10.756 & 99.80 & 10.998 & 100.0 & 11.000 & 100.0 & 11.000\\
& AIC & 99.7 & 10.997 & 100.0 & 11.000 & 100.0 & 11.000 & 100.0 &
11.000\\\hline
$T=1.5n$ & LY & 0.10 & 2.143 & 0.00 & 2.204 & 0.20 & 2.428 & 0.70 & 2.486\\
& BIC & 51.70 & 9.885 & 98.70 & 10.987 & 100.0 & 11.000 & 100.0 & 11.000\\
& HQIC & 97.80 & 10.973 & 100.0 & 11.000 & 100.0 & 11.000 & 100.0 & 11.000\\
& AIC & 100.0 & 11.000 & 100.0 & 11.000 & 100.0 & 11.000 & 100.0 &
11.000\\\hline
\multicolumn{10}{l}{See the notes of Table \ref{MC_table_1}.}\\\hline\hline
\end{tabular}
\end{center}
\end{table}

Finally, as suggested by a referee we build an alternative Monte Carlo design
that is based on the features of the data that we use in Section 4. Our
favorite DRVAR empirical specification is a model with $r=8$ and $p=2$ for a
dataset of $211$ aggregate US time series and $242$ observations. Obviously,
$n$ cannot be increased beyond $211$ in this framework but we can evaluate the
effects of doubling the sample size $T$ on the proposed methods. Hence, data
are generated by a DRVAR having Gaussian errors and the same parameters as the
model that we estimate in Section 4. The results, reported in \ref{MC_table_4}%
, confirm most of the main findings of the previous experiments, with the
exceptions that AIC outperforms BIC and that LY behaves even more poorly than
it does with the artificial DGP.\footnote{Again, the RFDs of FGLS are about
1\% lower than than those of OLS but the differences are significant in all
cases with the data-based DGP.}

\begin{table}[ptb]
\caption{Monte Carlo results, data-based DGP, OLS estimator}%
\label{MC_table_4}
\begin{center}%
\begin{tabular}
[c]{llcccccccc}\hline\hline
$n/T$ &  & \multicolumn{4}{c}{$T=242$} & \multicolumn{4}{c}{$T=484$%
}\\\cline{3-10}
&  & $\%\widehat{r}=8$ & $\%\widehat{r}<8$ & $\overline{\widehat{r}}$ & RFD &
$\%\widehat{r}=8$ & $\%\widehat{r}<8$ & $\overline{\widehat{r}}$ & RFD\\\hline
$n=211$ & LY & 0.0 & 100.0 & 1.735 & - & 1.8 & 98.2 & 1.816 & \\
& BIC & 8.8 & 91.2 & 6.464 & 78.93 & 55.0 & 45.0 & 7.548 & 55.97\\
& HQIC & 62.9 & 34.7 & 7.624 & 75.05 & 96.6 & 1.8 & 7.998 & 53.13\\
& AIC & 52.8 & 3.1 & 8.709 & 77.00 & 71.8 & 0.0 & 8.448 & 54.25\\\hline
\multicolumn{10}{l}{Notes: The DGP has the same parameters as the empirical
model in Section 4 and}\\
\multicolumn{10}{l}{Gaussian errors. See the notes of Table \ref{MC_table_1}
for the notation.}\\\hline\hline
\end{tabular}
\end{center}
\end{table}

Remarkably, the general outcome that (consistent) information criteria are
particularly useful in selection of VAR models is fully in line with previous
contributions in the literature (see e.g. Cavaliere et al., 2015, 2018;
Kapetanios, 2004; Nielsen, 2006). Regarding the poor performances of the LY
procedure, we remark that, differently from our DGPs, the dynamic component
and the white noise are generated independently in the Monte Carlo study by
Lam in Yao (2012). As noted by Lam \textit{et al.} (2011), the cross
correlation between the dynamic and the static component is beneficial to the
estimation of $A$, resulting even in a faster convergence rate as
$n^{-\delta/2}T^{1/2}$ when such cross correlation is strong, whereas it
creates difficulties in the estimation of $r$ by the LY procedure, see the
asymptotic analysis in Lam in Yao (2012).

\section{Empirical application}

This section illustrates the feasibility and the practical value of our
dimension reduction approach to VAR modelling. We first search for the
presence of co-movements among 211 US quarterly economic and financial time
series. Based on our methodology, we then propose a novel approach to identify
the shock that is responsible for most of the variability of the common
components at the business cycle frequencies.

\subsection{Co-movements in quarterly US time series}

We start by investigating a dimension reduction for a high-dimensional VAR of
the US economy. The data are obtained from the Federal Reserve Economic
Quaterely Database (FRED-QD henceforth), to which we added the total factor
productivity time series corrected for utilization produced by Fernald
(2012).\footnote{This variable is relevant for the structural analysis that we
conduct in the next subsection.} FRED-QD is regularly updated and different
releases of all the series are available online. A detailed description of the
variables and the proposed transformations used to achieve stationarity of
each series is provided by McCracken and Ng (2020).

After some necessary cleaning of the dataset and various stationarity
transformations of the series, we have at disposal $n=211$ variables with
$T=242$ quarterly observations from 1959Q3 up to 2019Q4.\footnote{We did not
include the variables that were not observed from 1959Q1 up to 2019Q4.} For
comparison convenience, we use the transformations proposed in FRED-QD to make
the time series stationary, although the results of unit root tests might
suggest alternative transformations in some cases. Moreover, series are
demeaned and standardized to have a unit variance after having corrected them
for outliers.\footnote{In particular, we removed 20 outliers from 16 series.
Our dataset is available upon request.}

We start the analysis by fixing $p_{0}=5$, a rather typical lag length of a
VAR model for quarterly data, and $R=14$ as the upper bound of the dimension
of the dynamic component.\footnote{The results that will be reported later are
robust to alternative choices as $p_{0}=3,...,9$ and $R=12,...,15.$} The LY
procedure detects $r=2$ dynamic components. This finding is rather dubious
given the huge heterogeneity in the series we work with. In order to determine
the largest VAR order, we use the traditional information criteria to estimate
the lag length in a VAR model for series $\hat{V}_{R}^{\prime}Y_{t}%
$.\footnote{Notice that, if series $Y_{t}$ are generated by a DVAR model, then
the linear combinations $V_{R}^{\prime}Y_{t}$ follow a VAR model of the same
order as the dynamic component $x_{t}$.} We get $p=1$ according to the BIC,
$p=2$ according to the HQIC, and $p=4$ according to the AIC. Consequently, we
consider successively $p=1,...,4$ lags when estimating $r$ through the
information criteria (\ref{IC}) using OLS and FGLS in estimation. As in the
Monte Carlo study, the two estimation methods provide identical estimates of
$r$: BIC [AIC] systematically indicates $r=7$ [$r=14$], whereas HQIC indicates
either $r=8$ or $r=12$ according to the VAR lag length. After a careful
comparison of the various specifications, we opt for $r=8$ and $p=2$, with the
latter being the indication coming from the HQIC when it is used to determine
the lag length having fixed $r=8$.\footnote{Choosing either $r=12$ or $p=1$
leads to empirical results of the subsequent analysis that are qualitatively
very similar to the ones that will be documented later.}\ We remark that the
presence of eight common components is a rather typical finding in the
empirical literature on factor models using similar data as ours.

Next, we compute two statistics in order to evaluate how the model fits to the
data. First, we consider the coefficients of determination of each element of
$Y_{t}$ as obtained by model (\ref{RR-MAI}). Second, we compute the squared
correlation coefficients between each element of $Y_{t}$ and its counterpart
in the common component $\chi_{t}$ of representation (\ref{Sdecomp}). We
denote the former statistic as $R_{Y,Z}^{2}$\ and the latter as $R_{Y,\Xi}%
^{2}$. It is easy to see that $R_{Y,\Xi}^{2}\geq$ $R_{Y,Z}^{2}$.

Whereas $R_{Y,Z}^{2}$ has the usual interpretation in terms of measure of the
degree of predictability, $R_{Y,\Xi}^{2}$ indicates the fraction of the
variability of each element of $Y_{t}$ that is explained by a linear
projection on the present and past values of the dynamic errors $\xi_{t}$.
Hence, $R_{Y,\Xi}^{2}$\ measures the importance of the common component
$\chi_{t}$ in the variability of each series.\footnote{Notice that $R_{Y,\Xi
}^{2}$ is actually computed as the sample analogous of the squared correlation
coefficient between each element of $Y_{t}$ and the corresponding element in
$Y_{t}-\nu_{t}$.}

Based on the FGLS estimates of the coefficients of the DRVAR model, we report
in Table \ref{Table_macro_2} the averages as well as the quartiles of the
empirical distributions of both $R_{Y,Z}^{2}$\ and $R_{Y,\Xi}^{2}$.

\begin{table}[th]
\caption{Average and quartiles of the measures of fit }%
\label{Table_macro_2}
\begin{center}%
\begin{tabular}
[c]{ccccc}\hline\hline
& Mean & $Q_{1}$ & $Q_{2}$ & $Q_{3}$\\\hline
$R_{Y,Z}^{2}$ & 0.30 & 0.12 & 0.25 & 0.47\\
$R_{Y,\Xi}^{2}$ & 0.53 & 0.34 & 0.56 & 0.73\\\hline
\end{tabular}
\end{center}
\end{table}

Moreover, in Table \ref{Table_macro_3} we report the estimates of both
$R_{Y,Z}^{2}$\ and $R_{Y,\Xi}^{2}$ for nine macroeconomic variables that we
are going to analyze in the subsequent subsection: Output (GDP), Consumption
(Con), Investment (Inv), Unemployment Rate, (UR), Worked Hours (Hours),
Inflation Rate (Inf), Interest Rate (IR), Labor Productivity (LP), and Total
Factor Productivity (TFP). The exact denominations of the variables along with
their stationarity transformations are reported in the appendix.

\begin{table}[th]
\caption{Measures of fit for 9 key aggregate variables}%
\label{Table_macro_3}
\begin{center}%
\begin{tabular}
[c]{cccccccccc}\hline\hline
& GDP & Con & Inv & UR & Hours & Inf & IR & LP & TFP\\\hline
$R_{Y,Z}^{2}$ & 0.40 & 0.38 & 0.40 & 0.64 & 0.61 & 0.22 & 0.28 & 0.20 & 0.10\\
$R_{Y,\Xi}^{2}$ & 0.83 & 0.71 & 0.73 & 0.89 & 0.85 & 0.90 & 0.68 & 0.66 &
0.37\\\hline
\end{tabular}
\end{center}
\end{table}

We see that, as expected, the estimates of $R_{Y,\Xi}^{2}$ are considerably
larger than those of $R_{Y,Z}^{2}$ over all the series and the nine key
variables as well. Remarkably, the role of the common component of TFP is
smaller than the one of the other key variables. This finding may reflect the
partial exogenous nature of TFP, as well as the possible presence of large
estimation errors in a variable that it is not directly observable.

\subsection{Comparison with a FAVAR model}

In light of Remark 4, it is of interest to check wether a FAVAR or an
(unrestricted)\ DRVAR fits better to the data. The first step of such
comparison requires to estimate a FAVAR model on our dataset. We use the nine
key variables as the observed factors and the remaining variables to construct
the unobserved factors, which are estimated by the principal components of the
remaining $202$ variables.

In order to fix the number of unobserved factors, rather that relying on
criteria that take into account the internal variability of the predictors
only, we follow Pesaran \textit{et al}. (2011) and use the traditional
information criteria in a predictive model where the target variables are the
nine key series and the predictors are two lags of both the targets and the
estimates of the unobserved factors The BIC and HQIC respectively suggest the
presence of $1$ and $3$ unobserved factors, whereas the AIC hits the upper
bound. As in the case of the DRVAR, we follow the indication coming from the HQIC.

In light of Proposition 1, it is not obvious how to rank the two empirical
models at the system level. A possible solution is contrasting the alternative
specifications of the partial model of the key series only (i.e., the
equations of the large VAR corresponding to the nine key variables). Given the
different number of parameters in the DRVAR and FAVAR specifications of the
considered partial model, a simple comparison of the respective
log-likelihoods would be misleading. Hence, Table \ref{Table_comparison}
reports the results of the application of traditional information criteria to
the competing specifications of the partial model for the key series, where
the estimated factors are treated as observable in the computation of the
penalty terms.\footnote{Since in both models the unobservable factors are
estimated by the eigenvectors of large covariance matrices, the sample
variability of the two estimators is expected to be similar.}

\begin{table}[th]
\caption{Information criteria of the DRVAR and FAVAR partial models of the key
series}%
\label{Table_comparison}
\begin{center}%
\begin{tabular}
[c]{cccc}\hline\hline
& BIC & HQIC & AIC\\\hline
DRVAR & -7.52 & -8.62 & -9.37\\
FAVAR & -6.62 & -8.48 & -9.73\\\hline
\end{tabular}
\end{center}
\end{table}

We see that BIC clearly supports the DRVAR specification, whereas AIC [HQIC]
favors the FAVAR [DRVAR]. Hence, notwithstanding the specification of the
FAVAR is tailored for the key variables, the two empirical models fit
similarly to the key series.

\subsection{Identification of the shock driving the business cycle}

A relevant issue in macroeconomics is the identifications of the shocks that
drive the macroeconomic fluctuations. The textbook approach consists in
identifying a given structural shock according to the guidance of economic
theory (e.g. a productivity shock or a monetary shock) and to evaluate its
impact over the key macroeconomic variables at various time horizons by means
of the impulse response function and the forecast error variance decomposition.

Recently, this strategy has been subject of some criticism on the ground that
empirical findings are strictly conditional on the validity of the underlying
economic assumption. For instance, it may be too restrictive to assume that
technology is the only shock that can permanently affect productivity.
Alternatively, several authors have resorted to identification schemes that
are based on the max-share identification strategy, as originally proposed by
Uhlig (2003). The max-share methodology is a kind of reverse engineering
approach, through which a shocks is identified as the main driver of a
macroeconomic variable at a given time horizon. Contributions along this line
of research include Barsky and Sims (2011) and Francis \textit{et al}. (2014).
Lately, Angeletos \textit{et al}. (2020) proposed a max-share approach to
identify the main driver of the business cycle. In particular, they identify
the main business cycle shock as the shock that maximizes the volatility at
the business cycle frequency band of a target variable. Using a Bayesian VAR
model for ten macroeconomic variables, they show that alternatively targeting
unemployment, output, hours worked, consumption and investment, their approach
provides very similar impulse responses functions for ten key macroeconomic variables.

The approach that we adopt here is similar as the one by Angeletos \textit{et
al}. (2020) but with some relevant differences. First, we rely on a much
richer information set coming from the large dimensional VAR that we have
previously estimated. Second, we recognize that the business cycle is
inherently a multivariate phenomenon and we aim at disentangling a unique
driver of the business cycle for the whole economy rather than targeting a
specific variable. Third, we search for the shock of the direction that
maximizes the variability at the business cycle frequencies of the common
component $\chi_{t}$ in decomposition (\ref{Sdecomp}). In this way, we are
able to filter out the effect of the ignorable errors, which cannot generate
cyclical fluctuations by construction but still contaminate the observed
variables $Y_{t}$. Notice that the last goal could not be pursued simply by
estimating a large VAR with some shrinkage method.

Formally, in view of Equation (\ref{Sdecomp}), the spectral density matrix of
the common component $\chi_{t}$ is%
\begin{equation}
F_{\varkappa}(\varpi)=(2\pi)^{-1}C(z^{-1})A^{\prime}\Sigma_{{\small u}%
}AC(z)^{\prime}, \label{Spec_Chi}%
\end{equation}
where $z=\exp(-i\varpi)$ and $\varpi\in\lbrack0,2\pi)$.

Given that $\operatorname{Re}F_{\varkappa}(\varpi)$ is proportional to the
variance matrix of the $\varpi$-frequency component in the spectral
representation of $\chi_{t}$\ (see, e.g., Subsections 4.6 and 7.1 in
Brillinger, 2001), the matrix
\[
\Theta(\varpi_{0},\varpi_{1})=\int_{\varpi_{0}}^{\varpi_{1}}\operatorname{Re}%
F_{\varkappa}(\varpi)\mathrm{d}\varpi
\]
measures the (co-)volatility of the common component $\chi_{t}$ at the
frequency band $[\varpi_{0},\varpi_{1}]$, where $0<\varpi_{0}<\varpi_{1}<\pi$.

Let $Q$ be the matrix formed by the eigenvectors that are associated with the
first $r$ non-increasing eigenvalues of the matrix $\Theta(\varpi_{0}%
,\varpi_{1})$, then the linear combinations $Q^{\prime}\chi_{t}$\ represent
the (static) principal components of $\chi_{t}$\ at the frequency band
$[\varpi_{0},\varpi_{1}]$. The Wold representation and the structural vector
moving average representation of $Q^{\prime}\chi_{t}$ are respectively given
by%
\[
Q^{\prime}\chi_{t}=Q(L)D\xi_{t}=Q^{\prime}\Psi(L)u_{t},
\]
where $D=Q^{\prime}C_{0}$, $Q(L)=Q^{\prime}C(L)D^{-1}$, $\Psi(L)=C(L)D^{-1}C$,
$u_{t}=C^{-1}D\xi_{t}$, and $C$ is a lower triangular matrix such that
$CC^{\prime}=DA^{\prime}\Sigma_{{\small u}}AD^{\prime}$.

When $[\varpi_{0},\varpi_{1}]$ is the typical business cycle frequency band,
i.e. $[2\pi/32,2\pi/6]$ for quarterly data, we label the first element in
$u_{t}$\ as the Main Business Cycle Common Shock (MBCCS). In words, the MBCCS
is the (standardized) shock of the direction that maximizes the
contemporaneous variability of the common component $\chi_{t}$ at frequencies
corresponding to periods between $6$ and $32$ quarters.\footnote{Notice that,
since we are considering the real part only of the spectral density matrix
$F_{\varkappa}(\varpi)$, we are implicitly focusing on the waves of $\chi_{t}$
at the frequency band $[\pi/16,\pi/3]$ that move in phase.} It is easy to see
that the MBCCS is
\[
u_{1t}=c_{11}^{-1}D_{1\bullet}^{\prime}\xi_{t}%
\]
where $c_{11}$ is the entry at the first row and first column of $C$ and
$D_{1\bullet}^{\prime}$\ is the first row of $D$, whereas, in light of
decomposition (\ref{Sdecomp}), the associated impulse response function (IRF)
for series $Y_{t}$ is%
\[
\Psi_{\bullet1}(L)=C(L)D^{-1}C_{\bullet1},
\]
where $C_{\bullet1}$ is the first column of $C$.

Given the frequency domain nature of our identification scheme, we evaluate
the effects of the MBCCS on the variable of interest over frequencies rather
than over time horizons (Centoni and Cubadda, 2003; Angeletos \textit{et al}.,
2020). In particular, we look at the contribution of the MBCCS to the
variability of the $i-$th series at the business cycle frequency band%
\[
\frac{\int_{\pi/16}^{\pi/3}e_{i}^{\prime}\Psi_{\bullet1}(z^{-1})\Psi
_{\bullet1}(z)^{\prime}e_{i}\mathrm{d}\varpi}{2\pi\int_{\pi/16}^{\pi/3}%
e_{i}^{\prime}F_{Y}(\varpi)e_{i}\mathrm{d}\varpi},
\]
and at the zero frequency%
\[
\frac{e_{i}^{\prime}\Psi_{\bullet1}(1)\Psi_{\bullet1}(1)^{\prime}e_{i}}{2\pi
e_{i}^{\prime}F_{Y}(0)e_{i}},
\]
where $e_{i}$ is an $n-$vector with unity as its $i-$th element and zeros
elsewhere, and the spectral density matrix of series $Y_{t}$ is
\[
F_{Y}(\varpi)=F_{\varkappa}(\varpi)+(2\pi)^{-1}\mathrm{E}(\nu_{t}\nu
_{t}^{\prime})
\]

Regarding the computational aspects, we truncate the order of the estimated
polynomial matrix $C(L)$ by setting $\hat{C}_{j}=0$ for $j>199$ and we
approximate integrals over the business cycle frequency band with sums over
$100$ evenly spaced frequencies from $\pi/16$ up to $\pi/3$.

\begin{figure}[ptb]%
\centering
\includegraphics[
trim=0.000000in 0.000000in -1.750069in 0.000000in,
height=3.5613in,
width=7.8663in
]%
{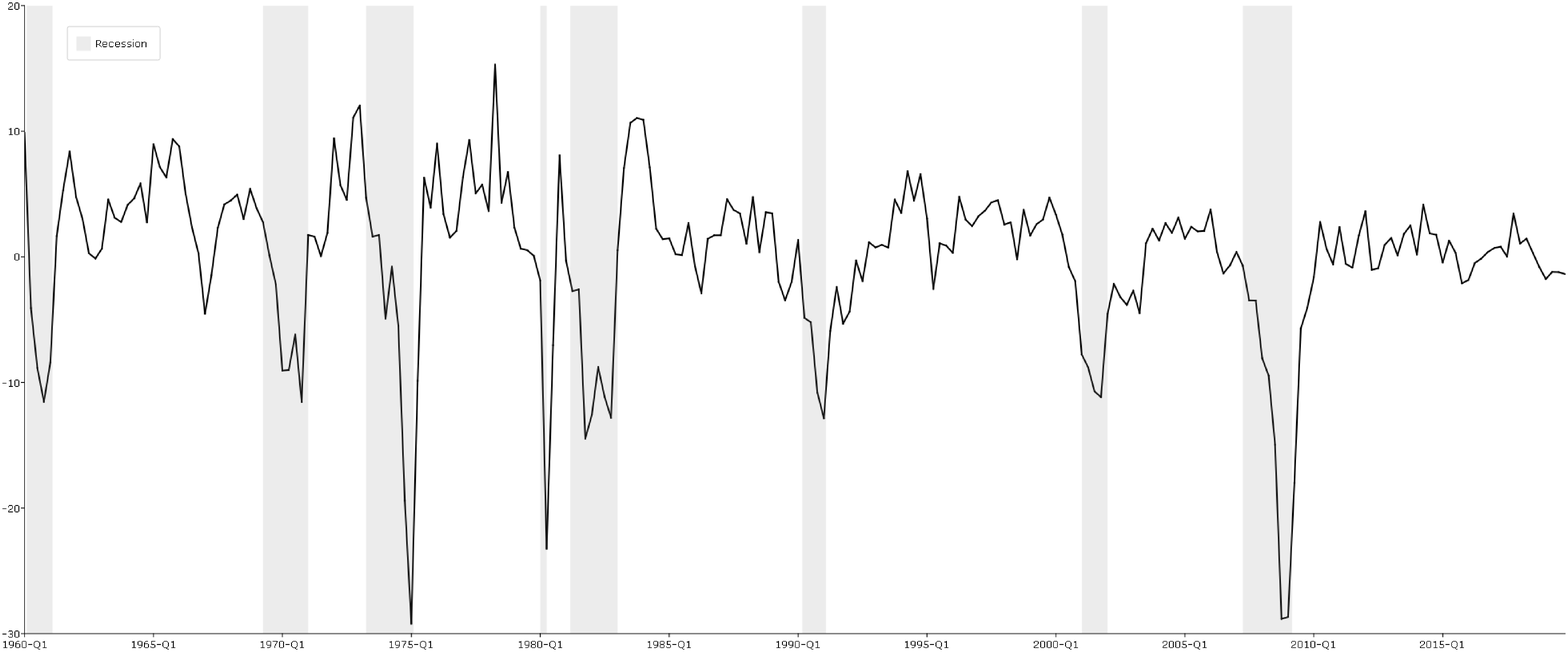}%
\caption{Estimate of the main business cycle common component }%
\label{PC1}%
\end{figure}

We start the empirical analysis by computing the sample eigenvalues of matrix
$\Theta(\pi/16,\pi/3)$. The largest eigenvalue, namely the one that is
associated with the MBCCS, accounts for about $54.8\%$ of the variability of
the common component at the business cycle frequency band. Figure (\ref{PC1})
shows the estimate of the linear combination of the common component that is
associated with the MBCCS, i.e. $Q_{1\bullet}^{\prime}\chi_{t}$\ where
$Q_{1\bullet}^{\prime}$ is the first row of $Q$, which we label as the Main
Business Cycle Common Component (MBCCC). Shaded areas in the figure represent
NBER-defined recessions. We see that the MBCCC accurately reproduces the main
features of the US aggregate cycle. Moreover, its estimated spectrum exhibits
a unique large peak corresponding to fluctuations with a period of about five years.%

%TCIMACRO{\FRAME{ftbpFU}{7.8663in}{3.5613in}{0pt}{\Qcb{Estimate of the main
%business cycle common component }}{\Qlb{PC1}}{pc1.eps}%
%width 7.8663in;  height 3.5613in;  depth 0pt;  original-width 16.479in;
%original-height 8.2192in;  cropleft "0";  croptop "1";  cropright "1.1062";
%cropbottom "0";  filename 'PC1.eps';file-properties "XNPEU";}} }%
%BeginExpansion
%EndExpansion

We report in Table \ref{Table_MBCCS} the estimates of the contributions of the
MBCCS to the variability of the 9 key series at periods 6-32 and $\infty$.
This information helps to asses the role of the MBCCS both at the business
cycle frequencies and in the long-run.

\begin{table}[th]
\caption{Contributions of the MBCCS to the variability of the 9 key series at
frequencies $[\pi/16,\pi/3]$ and 0}%
\label{Table_MBCCS}
\begin{center}%
\begin{tabular}
[c]{cccccccccc}\hline\hline
Period & GDP & Con & Inv & UR & Hours & Inf & IR & LP & TFP\\\hline
$6-32$ & $40.2$ & $15.3$ & $47.9$ & $44.2$ & $43.7$ & $8.7$ & $28.9$ & $21.0$
& $5.4$\\
$\infty$ & $33.3$ & $16.8$ & $35.8$ & $39.4$ & $37.7$ & $7.2$ & $19.4$ &
$14.4$ & $2.4$\\\hline\hline
\end{tabular}
\end{center}
\end{table}

\begin{figure}[ptb]%
\centering
\includegraphics[
trim=0.000000in 0.000000in -2.212496in 0.000000in,
height=3.5578in,
width=7.8646in
]%
{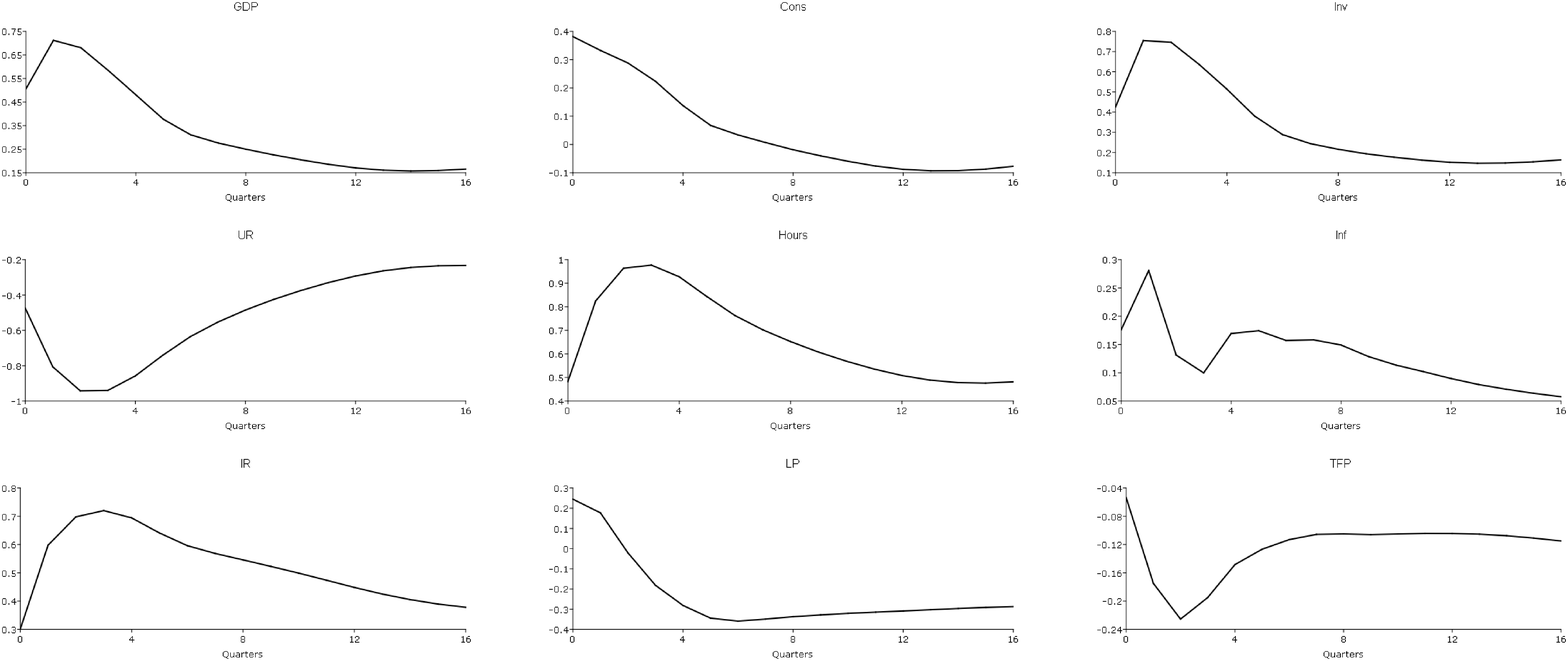}%
\caption{Cumulated impulse responses of the MBCCS}%
\label{IRF}%
\end{figure}

Finally, Figure (\ref{IRF}) reports the estimates of the cumulated IRF to the
MBCCS of the 9 variables already considered in the previous subsection. The
IRF are cumulated to provide readers with the dynamic effects of the MBCCS on
the levels of variables, thus facilitating the comparison with results of
previous studies.%

%TCIMACRO{\FRAME{ftbpFU}{7.8646in}{3.5578in}{0pt}{\Qcb{Cumulated impulse
%responses of the MBCCS}}{\Qlb{IRF}}{lvar_cirf.eps}%
%width 7.8646in;  height 3.5578in;  depth 0pt;  original-width 20.8333in;
%original-height 10.3855in;  cropleft "0";  croptop "1";  cropright "1.1062";
%cropbottom "0";  filename 'LVAR_cirf.eps';file-properties "XNPEU";}} }%
%BeginExpansion
%EndExpansion

We see in Figure (\ref{IRF}) that the MBCCS triggers a procyclical effect on
GDP and Inv peaking with one quarter delay, on Con with no delay, as well as
on UR [Hours and IR] with a peak at two [three] quarters. Moreover, in view of
Table \ref{Table_MBCCS}, it explains a large fraction of the cyclical
variability of Inv, Hours, UR and GDP.\footnote{In general, the contribution
of the MBCCS to the cyclical variability of the various variables is smaller
than those of the shocks that Angeletos \textit{et al}. (2020) obtain by
targeting UR or other procyclical variables. However, it should be reminded
that the aforementioned authors make use of a VAR with 10 series, whereas in
our empirical model each series is hit by 211 reduced form errors. This
considerably reduces the chances that a single shock can generate a very large
fraction of cyclical volatility.} These results corroborate the claim that the
considered shock is the main driver of the business cycle fluctuations.

However, the MBCCS has a limited positive impact on Inf, which peaks at one
quarter and slowly dies out, and it marginally affects both LP and,
especially, TFP. Moreover, it explains a small portion of the cyclical
movements of both Inf and TFP.

Regarding the long-run scenario, the MBCCS explains around 35\% of the
zero-frequency variability of UR, Hours, Inv, and GDP, whereas it has
negligible explanatory power for the permanent variation in Inf and TFP.
Surprisingly, the MBCCS is responsible for almost the same portion of
variability of Con (around 14\%) both in the long and short run.\footnote{With
the important caveat in mind that the validity of bootstrap for the DRVAR
needs to be formally proven, the contributions of the MBCCS to the variability
of GDP, INV, UR, HOURS at periods 6-32 and $\infty$ are larger than twice
their bootstrap standard errors. For the same variables and IR, some short
horizons of their IRFs are significant at the the 68\% level. Bootstrap was
implemented by sampling with replacement from the DRVAR residuals.}

All in all, the findings above seem to preclude the interpretation of the
MBCCS as either a productivity or a news shock on the one hand, and a
traditional demand shock on the other hand. However, a meticulous
interpretation of the empirical results of this application, in particular by
means of a rigorous comparison with DSGE models, is beyond the scope of the
present paper.

\section{Conclusions}

This paper provides a link between two related but different strands of the
literature on dimension reduction of multivariate time series, namely dynamic
factor modelling on the one hand and the common feature methodology on the
other hand. The former approach has the advantage that a limited number of
factors is often enough to summarize the variation of a large dataset in
economic and financial applications. However, the common feature approach has
the nice theoretical property that the related common components posses a
given time series feature (autocorrelation, volatility, trends, etc.) whereas
the uncommon components do not.

Building on Lam \textit{et al.} (2011) and Lam and Yao (2012), we propose a
dimensional reduction approach such that both a common right space and a
common left null space are present in the coefficient matrices of a large VAR
model. This specification allows to detect a small dimensional VAR that is
responsible for the whole dynamics of the system. This approach has many
potential applications such as forecasting big data from a small scale VAR
without loosing relevant information, structural VAR analysis, realized
covariance matrices modelling, etc.

Our Monte Carlo study shows that we should consider either the BIC or the HQIC
to detect the number of the common dynamic components. We illustrate the
feasibility of our framework on large dimensional macroeconomic times series
relative to the US economy. Around 8 common components generate the entire
dynamics of 211 aggregate economic variables. Moreover, we offer a novel
approach to identify the shock that is responsible for most of the volatility
of the common component at the business cycle frequency band. Such shock as a
clear expansionary effect, both in the short and long run, on the labor market
variables, output, and investments but it affects marginally inflation and
total factor productivity.

\section{Appendix}

In this appendix we prove the main results of Subsection 2.2. and provide the
details about the key variables that are analyzed in Section 4.

\subsection{Mean square errors of the OLS and GLS estimators}

In this subsection we provide the proof of Theorem 6. The thesis is given by
the following inequality:%
\begin{equation}
\mathrm{E}[\mathrm{Vec}(\hat{\alpha}-\alpha)\mathrm{Vec}(\hat{\alpha}%
-\alpha)^{\prime}|Z]\geq\mathrm{E}[\mathrm{Vec}(\tilde{\alpha}-\alpha
)\mathrm{Vec}(\tilde{\alpha}-\alpha)^{\prime}|Z] \label{Ineq}%
\end{equation}
for\ $\forall\mathrm{Vec}(\alpha)\in%
%TCIMACRO{\U{211d} }%
%BeginExpansion
\mathbb{R}
%EndExpansion
^{nr}$.

Let us start with inserting Equation (\ref{Y}) into (\ref{Vec(alpha_hat)}) and
(\ref{Vec(alpha_tilde)})\ to respectively get%
\begin{equation}
\mathrm{Vec}(\hat{\alpha})=\mathrm{Vec}(\alpha)+[A^{\prime}\otimes(Z^{\prime
}Z)^{-1}Z^{\prime}]\mathrm{Vec}({\small u}) \label{alpha_hat}%
\end{equation}%
\begin{equation}
\mathrm{Vec}(\tilde{\alpha})=\mathrm{Vec}(\alpha)+\left[  \left(  A^{\prime
}\Sigma_{{\small u}}^{-1}A\right)  ^{-1}A^{\prime}\Sigma_{{\small u}}%
^{-1}\otimes\left(  Z^{\prime}Z\right)  ^{-1}Z^{\prime}\right]  \mathrm{Vec}%
({\small u}) \label{alpha_tilde}%
\end{equation}

In view of Equations (\ref{alpha_hat}) and (\ref{alpha_tilde}) we get
\[
\mathrm{E}[\mathrm{Vec}(\hat{\alpha}-\alpha)\mathrm{Vec}(\hat{\alpha}%
-\alpha)^{\prime}|Z]=A^{\prime}\Sigma_{{\small u}}A\otimes\left(  Z^{\prime
}Z\right)  ^{-1}%
\]
and
\[
\mathrm{E}[\mathrm{Vec}(\tilde{\alpha}-\alpha)\mathrm{Vec}(\tilde{\alpha
}-\alpha)^{\prime}|Z]=\left(  A^{\prime}\Sigma_{{\small u}}^{-1}A\right)
^{-1}\otimes\left(  Z^{\prime}Z\right)  ^{-1}%
\]

Since the Kronecker product of two positive semidefinite matrix is positive
semidefinite as well, proving (\ref{Ineq}) requires to show that
\begin{equation}
A^{\prime}\Sigma_{{\small u}}A\geq\left(  A^{\prime}\Sigma_{{\small u}}%
^{-1}A\right)  ^{-1} \label{Ineq_2}%
\end{equation}
In order to prove inequality (\ref{Ineq_2}), we notice that%
\[
B^{\prime}\Sigma_{{\small u}}B=(B^{\prime}\Sigma_{{\small u}}^{-1}B)^{-1}%
\]
where $B=[A,A_{\perp}]$. Partitioning conformably in blocks both sides of the
equation above we get%
\begin{equation}
\left[
\begin{array}
[c]{cc}%
A^{\prime}\Sigma_{{\small u}}A & A^{\prime}\Sigma_{{\small u}}A_{\perp}\\
A_{\perp}^{\prime}\Sigma_{{\small u}}A & A_{\perp}^{\prime}\Sigma_{{\small u}%
}A_{\perp}%
\end{array}
\right]  =\left[
\begin{array}
[c]{cc}%
A^{\prime}\Sigma_{{\small u}}^{-1}A & A^{\prime}\Sigma_{{\small u}}%
^{-1}A_{\perp}\\
A_{\perp}^{\prime}\Sigma_{{\small u}}^{-1}A & A_{\perp}^{\prime}%
\Sigma_{{\small u}}^{-1}A_{\perp}%
\end{array}
\right]  ^{-1} \label{blocks}%
\end{equation}
Applying the rule of the inverse of a partitioned symmetric matrix to the
upper left block of the matrices in (\ref{blocks}) we see%
\begin{align}
A^{\prime}\Sigma_{{\small u}}A  &  =(A^{\prime}\Sigma_{{\small u}}^{-1}%
A)^{-1}+\label{Upper_left}\\
&  (A^{\prime}\Sigma_{{\small u}}^{-1}A)^{-1}A^{\prime}\Sigma_{{\small u}%
}^{-1}A_{\perp}[A_{\perp}^{\prime}\Sigma_{{\small u}}^{-1}A_{\perp}-A_{\perp
}^{\prime}\Sigma_{{\small u}}^{-1}A(A^{\prime}\Sigma_{{\small u}}^{-1}%
A)^{-1}A^{\prime}\Sigma_{{\small u}}^{-1}A_{\perp}]^{-1}A_{\perp}^{\prime
}\Sigma_{{\small u}}^{-1}A(A^{\prime}\Sigma_{{\small u}}^{-1}A)^{-1}\nonumber
\end{align}
Finally, noticing that the matrix in square brackets in (\ref{Upper_left}) is
the conditional variance matrix of $A_{\bot}^{\prime}\Sigma_{{\small u}}%
^{-1}{\small u}_{t}$\ given $A^{\prime}\Sigma_{{\small u}}^{-1}{\small u}_{t}%
$, then inequality (\ref{Ineq_2}) trivially follows.

Moreover, the equality sign holds when $A^{\prime}\Sigma_{{\small u}}%
^{-1}A_{\perp}=0$, that is when the matrix $A$ belongs to the space spanned by
$r$ of the eigenvectors of $\Sigma_{{\small u}}^{-1}$, and consequently
$A_{\perp}$ belongs to the space spanned by the remaining $n-r$ eigenvectors.
Since the eigenvectors of $\Sigma_{{\small u}}^{-1}$ are the same of
$\Sigma_{{\small u}}$, we notice that that two estimators have the same
variance when $A^{\prime}{\small u}_{t}$ and $A_{\bot}^{\prime}{\small u}_{t}$
are not correlated.

\subsection{Consistency of the BIC and HQIC for $r$.}

In this subsection we show that the BIC and HQIC provide weakly consistent
estimators for the number of dynamic components $r$ but not for the overall
number of DRVAR parameters $k$. We start by reviewing the conditions on the
penalty term $c_{T}$ for weak consistency of an information criterion such as
(\ref{IC}) when $n$ is fixed and $T$ diverges: $c_{T}\rightarrow\infty$ and
$\frac{c_{T}}{T}\rightarrow0$ as $T\rightarrow\infty$, see e.g. L\"{u}tkepohl
(2005).\ It easily follows that BIC and HQIC are consistent for $k$ whereas
the AIC is not when $n$ is finite.

Let us assume that both OLS and FGLS estimate\ the DRVAR parameters (up to an
orthonormal transformation) with the standard $\sqrt{T}$ rate as
$n,T\rightarrow\infty$, and that $\gamma=\underset{n\rightarrow\infty}{\lim
}\left(
%TCIMACRO{\tprod \limits_{i=1}^{n}}%
%BeginExpansion
{\textstyle\prod\limits_{i=1}^{n}}
%EndExpansion
\sigma_{i}^{2}\right)  ^{1/n}$ exists. When $q\geq r$, $\hat{\sigma}_{i}%
^{2}(q)$\ is a $\sqrt{T}$-consistent estimator of $\sigma_{i}^{2}$ for any
$i$. When $q<r$, $\underset{n,T\rightarrow\infty}{\mathrm{plim}}\hat{\sigma
}_{i}^{2}(q)\geq\sigma_{i}^{2}$ for any $i$ and, given that the factors are
assumed to be strong, the case where each element of $\bar{a}_{j}$ is $O(1)$
for $j=1,...,r$\ is included, implying that the dynamic component $x_{t}$
influences most of the $n$ time series. Hence, it follows that%
\[
\underset{n,T\rightarrow\infty}{\mathrm{plim}}\hat{\gamma}_{q}=\left\{
\begin{array}
[c]{cc}%
\gamma & \text{for }q\geq r\\
\gamma^{\ast}>\gamma & \text{for }q<r
\end{array}
\right.
\]
where $\hat{\gamma}_{q}=\left(
%TCIMACRO{\tprod \limits_{i=1}^{n}}%
%BeginExpansion
{\textstyle\prod\limits_{i=1}^{n}}
%EndExpansion
\hat{\sigma}_{i}^{2}\right)  ^{1/n}$.

Let us rewrite Equation (\ref{IC}) as
\[
\mathrm{IC}(q)=\ln(\hat{\gamma}_{q})+\frac{c_{T}k}{Tn}%
\]
Since
\[
\lim_{n,T\rightarrow\infty}\frac{k}{n}=\underset{n\rightarrow\infty}{\lim
}\frac{nq+(p-1)q^{2}}{n}=q
\]
we conclude that $\mathrm{IC}(q)$ is asymptotically equivalent to
\[
\mathrm{IC}^{\ast}(q)=\ln(\hat{\gamma}_{q})+\frac{c_{T}}{T}q,
\]
and that $\mathrm{IC}^{\ast}(q)$\ is weakly consistent for $r$ (but not for
$k$) when the penalty term $c_{T}$\ is the one of the HQIC or BIC.

\subsection{Key variables and their transformations}

We report in Table \ref{Variables} the denominations of the nine key variables
that are analyzed in Section 4 along with the relative transformations.

\begin{table}[th]
\caption{Variables and their transformations}%
\label{Variables}
\begin{center}%
\begin{tabular}
[c]{lll}\hline\hline
Variable & Transformation & Acronym\\\hline
Real Gross Domestic Product & $(1-L)\log$ & GDP\\
Real Personal Consumption Expenditures & $(1-L)\log$ & Con\\
Real Gross Private Domestic Investment & $(1-L)\log$ & Inv\\
Civilian Unemployment Rate & $(1-L)$ & UR\\
Nonfarm Business Sector: Hours of All Persons & $(1-L)\log$ & Hours\\
Personal Consumption Expenditures: Chain-type Price Index & $(1-L)^{2}\log$ &
Inf\\
Effective Federal Funds & $(1-L)$ & IR\\
Nonfarm Business Sector: Real Output Per Hour of All Persons & $(1-L)\log$ &
LP\\
Total Factor Productivity & $(1-L)\log$ & TFP\\\hline
\end{tabular}
\end{center}
\end{table}


\begin{thebibliography}{99}                                                                                               %


\bibitem {}\textsc{Ahn, S. K. (1997)}, Inference of Vector Autoregressive
Models With Cointegration and Scalar Components, \emph{Journal of the American
Statistical Association}, 92, 350-356.

\bibitem {}\textsc{Ahn, S. K. and G. C. Reinsel (1988)}, Nested Reduced Rank
Autoregressive Models for Multiple Time Series, \emph{Journal of the American
Statistical Association}, 83, 849-856.

\bibitem {}\textsc{Angeletos, G.M., Collard, F., and Dellas, H. (2020),}
Business-Cycle Anatomy, \emph{American Economic Review}, 10, 3030-3070.

\bibitem {}\textsc{Athanasopoulos, G., Guillen, O.T., Issler, J.V., and Vahid,
F. (2011),} Model Selection, Estimation and Forecasting in VAR Models with
Short-run and Long-run Restrictions,\ \emph{Journal of Econometrics,} 164, 116-129.

\bibitem {}\textsc{Barsky, R. B. and Sims, E. R. (2011),} News shocks and
business cycles, \emph{Journal of Monetary Economics}, 58, 273-289.

\bibitem {}\textsc{Bernanke, B., Boivin, J., and P.S. Eliasz (2005),}
Measuring the Effects of Monetary Policy: A Factor-augmented Vector
Autoregressive (FAVAR) Approach, \emph{The Quarterly Journal of Economics},
120, 387-422.

\bibitem {}\textsc{Bernardini E. and G. Cubadda (2015),} Macroeconomic
Forecasting and Structural Analysis through Regularized Reduced-Rank
Regression, \emph{International Journal of Forecasting}, 31, 682-691.

\bibitem {}\textsc{Boivin, J. and S. Ng (2006)}, Are more data always better
for factor analysis, \emph{Journal of Econometrics}, 132, 169--194.

\bibitem {}\textsc{Brillinger, D. (2001),} \emph{Time series: data analysis
and theory}, Society for Industrial and Applied Mathematics.

\bibitem {}\textsc{Carriero, A. Kapetanios, G., and M. Marcellino (2011),}
Forecasting Large Datasets with Bayesian Reduced Rank Multivariate Models,
\emph{Journal of Applied Econometrics}, 26, 735-761.

\bibitem {}\textsc{Carriero, A. Kapetanios, G., and M. Marcellino (2016),}
Structural analysis with Multivariate Autoregressive Index models,
\emph{Journal of Econometrics}, 192, 332--348.

\bibitem {}\textsc{Cavaliere, G., L. De Angelis, A. Rahbek, and A.M.R. Taylor
(2015),} A comparison of sequential and information-based methods for
determining the co-integration rank in heteroskedastic VAR models,
\emph{Oxford Bulletin of Economics and Statistics}, 77, 106--128.

\bibitem {}\textsc{Cavaliere, G., L. De Angelis, A. Rahbek, and A.M.R. Taylor
(2018),} Determining the cointegration rank in heteroskedastic VAR models of
unknown order, \emph{Econometric Theory}, 34, 349--382.

\bibitem {}\textsc{Centoni, M., and G. Cubadda (2003),} Measuring the business
cycle effects of permanent and transitory shocks in cointegrated time series.
\emph{Economics Letters} 80, 45--51.

\bibitem {}\textsc{Centoni M., and G. Cubadda (2015),} Common feature analysis
of economic time series: An overview and recent developments,
\emph{Communications for Statistical Applications and Methods,} 22, 1--20.

\bibitem {}\textsc{Chamberlain, G., and M. Rothschild (1983)}, Arbitrage,
Factor Structure, and Mean-Variance Analysis on Large Asset Markets,
\emph{Econometrica}, 51, 1281-1304.

\bibitem {}\textsc{Chang, J. Y., Guo, B., and Q. Yao, (2018), }Principal
Component Analysis for Second-Order Stationary Vector Time Series,
\emph{Annals of Statistics}, 46, 2094--2124

\bibitem {}\textsc{Cubadda, G. (2007), }A Unifying Framework for Analyzing
Common Cyclical Features in Cointegrated Time Series, \emph{Computational
Statistics and Data Analysis,} 52, 896--906.

\bibitem {}\textsc{Cubadda, G., and B. Guardabascio (2019)}, Representation,
Estimation and Forecasting of the Multivariate Index-Augmented Autoregressive
Model, \emph{International Journal of Forecasting}, 35, 67--79.

\bibitem {}\textsc{Cubadda G., Guardabascio B., and A. Hecq (2017)}, A Vector
Heterogeneous Autoregressive Index Model for Realized Volatility Measures,
\emph{International Journal of Forecasting}, 33, 337--344.

\bibitem {}\textsc{Cubadda, G. and A. Hecq (2001),} On non-contemporaneous
short-run comovements, \emph{Economics Letters,} 73, 389-397.

\bibitem {}\textsc{Cubadda, G. and A. Hecq (2011),} Testing for Common
Autocorrelation in Data Rich Environments, \emph{Journal of Forecasting}, 30, 325--335.

\bibitem {}\textsc{Cubadda, G., and A. Hecq (2021),} Reduced Rank Regression
Models in Economics and Finance, \emph{Oxford Research Encyclopedia of
Economics and Finance}, Oxford University Press, doi:
10.1093/acrefore/9780190625979.013.677, forthcoming.

\bibitem {}\textsc{Cubadda, G., Hecq A. and F.C. Palm (2009),} Studying
Co-movements in Large Multivariate Models Prior to Modeling, \emph{Journal of
Econometrics,} 148, 25-35.

\bibitem {}\textsc{Davis, R.A., Zang, P. and T. Zheng (2016)}, Sparse Vector
Autoregressive Modeling, \emph{Journal of Computational and Graphical
Statistics}, 25, 1077--1096.

\bibitem {}\textsc{Engle, R.F., and S. Kozicki (1993),} Testing for Common
Features (with comments), \emph{Journal of Business and Economic Statistics},
11, 369--395.

\bibitem {}\textsc{Fernald, J.G. (2012),} A Quarterly, Utilization-adjusted
Series on Total Factor Productivity, \emph{Federal Reserve Bank of San
Francisco Working Paper Series}, 2012-19.

\bibitem {}\textsc{Fern\'{a}ndez-Villaverde, J., Rubio-Ram\'{\i}rez, J.F., and
F.Schorfheide (2016),} Solution and Estimation Methods For DSGE Models, in
Taylor, J.B. and H. Uhlig (eds), \emph{Handbook of Macroeconomics}, 2, 527-724.

\bibitem {}\textsc{Forni, M., Giannone, D., Lippi, M, and L. Reichlin (2009),}
Opening the black box: structural factor models with large cross sections,
\emph{Econometric Theory}, 25, 1319--1347.

\bibitem {}\textsc{Francis, N., Owyang, M. T., Roush, J. E., and Dicecio, R.
(2014),} A Flexible Finite-Horizon Alternative to Long-run restrictions with
an application to technology shocks, \emph{Review of Economic Statistics}, 96, 638-647.

\bibitem {}\textsc{Goetz, T., Hecq, A. and S. Smeekes (2016), }Testing for
Granger-causality in large mixed-frequency VARs, \emph{Journal of
Econometrics,} 193, (2), 418-432.

\bibitem {}\textsc{Hecq, A., Palm, F.C. and J.P. Urbain (2006),} Common
cyclical features analysis in VAR models with cointegration, \emph{Journal of
Econometrics,} 132, 117--141.

\bibitem {}\textsc{Hecq, A., Margaritella, L. and S. Smeekes (2021),} Granger
Causality Testing in High-Dimensional VARs: a Post-Double-Selection Procedure,
\textit{Journal of Financial Econometrics}, forthcoming.

\bibitem {}\textsc{Hsu, N. J. and H. L. Hung and Y. M. Chang (2008)}, Subset
selection for vector autoregressive processes using Lasso, \emph{Computational
Statistics \& Data Analysis, 52}, 3645-3657.

\bibitem {}\textsc{Hu, Z., Dong, K., Dai, W. and T. Tong (2017),} A Comparison
of Methods for Estimating the Determinant of High-Dimensional Covariance
Matrix, \emph{International Journal of Biostatistics,} 13, 1-24.

\bibitem {}\textsc{Kapetanios, G. (2004),} The asymptotic distribution of the
cointegration rank estimator under the Akaike information criterion,
\emph{Econometric Theory}, 20, 735--742.

\bibitem {}\textsc{Karlsson, S. (2013),} Forecasting with Bayesian Vector
Autoregression, \emph{Handbook of Economic Forecasting,} 2B, 791-897, North Holland.

\bibitem {}\textsc{Kock, A. B. and Callot, L. (2015),} Oracle inequalities for
high dimensional vector autoregressions, \emph{Journal of Econometrics,} 186, 325--344.

\bibitem {}\textsc{Kohn, R. (1982),} When is an aggregate of a time series
efficiently forecast by its past?, \emph{Journal of Econometrics}, 18, 337-349.

\bibitem {}\textsc{Koop, G. (2017),} Bayesian Methods for Empirical
Macroeconomics with Big Data, \emph{Review of Economic Analysis}, 9 , 33-56.

\bibitem {}\textsc{Lam, C. and Q. Yao (2012),} Factor modeling for
high-dimensional time series: Inference for the number of factors,
\emph{Annals of Statistics,} 40, 694-726.

\bibitem {}\textsc{Lam, C., Yao, Q., and N. Bathia, (2011),} Estimation of
latent factors for high-dimensional time series, \emph{Biometrika,} 98 901--918.

\bibitem {}\textsc{Li, W., Gao, J., Li K., and Q. Yao (2016),} Modeling
Multivariate Volatilities via Latent Common Factors, \emph{Journal of Business
\& Economic Statistics,} 34, 564-573.

\bibitem {}\textsc{Li, Z., Wang, Q.W., and J. Yao (2017), }Identifying the
number of factors from singular values of a large sample auto-covariance
matrix, \emph{Annals of Statistics,} 45, 257--288.

\bibitem {}\textsc{Lippi, M. (2018),} Frequency-Domain Approach in
High-Dimensional Dynamic Factor Models, \emph{Oxford Research Encyclopedia of
Economics and Finance}, Oxford University Press, doi: 10.1093/acrefore/9780190625979.013.171.

\bibitem {}\textsc{Lippi, M. (2019),} Time-Domain Approach in High-Dimensional
Dynamic Factor Models, \emph{Oxford Research Encyclopedia of Economics and
Finance}, Oxford University Press, doi: 10.1093/acrefore/9780190625979.013.169.

\bibitem {}\textsc{L\"{u}tkepohl, H (1984a), }Forecasting Contemporaneously
Aggregated Vector ARMA Processes, \emph{Journal of Business and Economic
Statistics}, 2, 201-214.

\bibitem {}\textsc{L\"{u}tkepohl, H (1984b), }Linear transformations of vector
ARMA processes, \emph{Journal of Econometrics}, 26, 283-293.

\bibitem {}\textsc{L\"{u}tkepohl, H (2005),} \emph{New Introduction to
Multiple Time Series Analysis}, Springer-Verlag Berlin.

\bibitem {}\textsc{McCracken, M.W. and Ng, S., (2020),} FRED-QD: A Quarterly
Database for Macroeconomic Research, \emph{Federal Reserve Bank of St. Louis
Working Paper}, 2020-005, https://doi.org/10.20955/wp.2020.005.

\bibitem {}\textsc{Nicholson, W., Wilms, I., Bien, J., and D.S. Matteson
(2018)}, High Dimensional Forecasting via Interpretable Vector Autoregression,
\emph{arXiv:1412.5250v3}.

\bibitem {}\textsc{Nielsen, B. (2006),} Order determination in general vector
autoregressions, \emph{IMS Lecture Notes - Monograph Series}, 52, 93--112.

\bibitem {}\textsc{Oberhofer, W. and J. Kmenta (1974),} A general procedure
for obtaining maximum likelihood estimates in generalized regression models,
\emph{Econometrica,} 42, 579--590.

\bibitem {}\textsc{Pesaran, M.H., Pick, A., and A. Timmermann (2011),}
Variable selection, estimation and inference for multi-period forecasting
problems, \emph{Journal of Econometrics}, 164, 173--187.

\bibitem {}\textsc{Quenouille, M.H. (1957)}, \emph{The Analysis of Multiple
Time Series}, Griffin's Statistical Monographs \& courses.

\bibitem {}\textsc{Reinsel, G., (1983)}, Some results on multivariate
autoregressive index models, \emph{Biometrika}, 70 (1), 145--156.

\bibitem {}\textsc{Sims, C. (1980),} Macroeconomics and Reality,
\emph{Econometrica}, 48, 1-48.

\bibitem {}\textsc{Smeekes, S. and E. Wijler, E. (2018),} Macroeconomic
forecasting using penalized regression methods, \emph{International Journal of
Forecasting,} 34, 408-430.

\bibitem {}\textsc{Stock, J. H. and Watson, M. W. (2016),} Dynamic Factor
Models, Factor-Augmented Vector Autoregressions, and Structural Vector
Autoregressions in Macroeconomics, in Taylor, J.B. and H. Uhlig (eds),
\emph{Handbook of Macroeconomics,} Vol. 2, North Holland.

\bibitem {}\textsc{Tao, M., Wang, Y, Yao, Q., and J Zou (2011),} Large
Volatility Matrix Inference via Combining Low-Frequency and High-Frequency
Approaches, \emph{Journal of the American Statistical Association, 106,} 1025-1040.

\bibitem {}\textsc{Uhlig, H. (2003),} What moves real GNP? \emph{mimeo HU
Berlin}.

\bibitem {}\textsc{Vahid, F., and R.F. Engle (1993)}, Common trends and common
cycles, \emph{Journal of Applied Econometrics}, 8, 341--360.

\bibitem {}\textsc{Velu, R., Reinsel, G., and D. Wichern (1986),} Reduced Rank
Models for Multiple Time Series, \emph{Biometrika}, 73, 105--118.

\bibitem {}\textsc{Wilms, I. and C. Croux (2016),} Forecasting using sparse
cointegration, \emph{International Journal of Forecasting}, 32(4), 1256-1267.

\bibitem {}\textsc{Zhang, R., Robinson, P., and Q. Yao (2019),} Identifying
Cointegration by Eigenanalysis, \emph{Journal of the American Statistical
Association,} 114, 916-927.
\end{thebibliography}
\end{document}